\documentclass[aps,prl,amsmath,twocolumns,amssymb,superscriptaddress,amsfonts,reprint]{revtex4-1}
\usepackage{graphicx}
\usepackage{bm,upgreek}
\usepackage{bm}

\usepackage[colorlinks=true,linkcolor=blue,citecolor=blue,urlcolor=blue]{hyperref}

\begin{document}



\title{Arrested coalescence of multicellular aggregates}

\author{David Oriola}\email{david.oriola@embl.es}
\affiliation{ EMBL Barcelona, Dr. Aiguader, 88, 08003, Barcelona, Spain}
\author{Miquel Marin-Riera}
\affiliation{ EMBL Barcelona, Dr. Aiguader, 88, 08003, Barcelona, Spain}
\author{Kerim Anlas}
\affiliation{ EMBL Barcelona, Dr. Aiguader, 88, 08003, Barcelona, Spain}
\author{Nicola Gritti}
\affiliation{ EMBL Barcelona, Dr. Aiguader, 88, 08003, Barcelona, Spain}
\author{Marina Matsumiya}
\affiliation{ EMBL Barcelona, Dr. Aiguader, 88, 08003, Barcelona, Spain}
\author{Germaine Aalderink}
\affiliation{ EMBL Barcelona, Dr. Aiguader, 88, 08003, Barcelona, Spain}
\author{Miki Ebisuya} 
\affiliation{ EMBL Barcelona, Dr. Aiguader, 88, 08003, Barcelona, Spain}
\author{James Sharpe} 
\affiliation{ EMBL Barcelona, Dr. Aiguader, 88, 08003, Barcelona, Spain}
\affiliation{Instituci\'o Catalana de Recerca i Estudis Avan\c{c}ats, 08010, Barcelona, Spain}
\author{Vikas Trivedi} \email{trivedi@embl.es} 
\affiliation{ EMBL Barcelona, Dr. Aiguader, 88, 08003, Barcelona, Spain}
\affiliation{EMBL Heidelberg, Developmental Biology Unit, 69117 Heidelberg, Germany}

\maketitle

{\bf Multicellular aggregates are known to exhibit liquid-like properties. The fusion process of two cell aggregates is commonly studied as the coalescence of two viscous drops. However, tissues are complex materials and can exhibit viscoelastic behaviour. It is known that elastic effects can prevent the complete fusion of two drops, a phenomenon known as arrested coalescence. Here we report the presence of this phenomenon in stem cell aggregates and provide a theoretical framework which agrees with the experiments. In addition, agent-based simulations show that cell protrusion activity controls a solid-to-fluid phase transition, revealing that arrested coalescence can be found in the vicinity of an unjamming transition. By analysing the dynamics of the fusion process and combining it with nanoindentation measurements, we obtain the effective viscosity, shear modulus and surface tension of the aggregates. More generally, our work provides a simple, fast and inexpensive method to characterize the mechanical properties of viscoelastic materials. } \\

Shaping of organs during morphogenesis results from the material response of the constituent tissues to the forces which in turn are generated by them. Understanding the material properties of biological tissues holds key to elucidate how shape and form emerge during morphogenesis both {\it in vivo} during embryonic development \cite{heisenberg2013forces,hamada2015role}, as well as {\it in vitro} in the context of synthetic morphogenesis \cite{teague2016synthetic, gritti2020rethinking, garreta2020rethinking}. For instance, viscous dissipation allows tissues to gradually change their shape without accumulation of significant stresses \cite{wyatt2016question,clement2017viscoelastic} and adapt to different environments. Embryonic tissues are known to exhibit liquid-like properties: they round up \cite{schotz2008quantitative,yu2018coherent}, fuse \cite{gordon1972rheological}, engulf other tissues \cite{steinberg1994experimental} and segregate or sort from heterotypic cell mixtures \cite{heintzelman1978liquid,foty1994liquid}. However, tissues are also known to exhibit elastic behaviour which can critically affect the final tissue configuration \cite{schotz2008quantitative,yu2018coherent,mongera2018fluid}. 
Unlike viscous forces, which only affect the rate of deformation of the tissue, elastic forces can resist deformation leading to a non-trivial final tissue configuration. Indeed jamming \cite{mongera2018fluid} and viscoelastic \cite{schotz2008quantitative,luu2011large} effects have been shown to be critical in different morphogenetic processes. \\

The mechanical properties of tissues have been measured using a wide range of techniques (for a detailed review see Refs. \cite{sugimura2016measuring,campas2016toolbox}). Absolute measurements of tissue mechanical parameters such as surface tension $\gamma$, viscosity $\eta$ or shear modulus $\mu$, are possible by means of different techniques such as parallel plate compression \cite{foty1996surface,schotz2008quantitative,forgacs1998viscoelastic}, axisymmetric drop shape analysis \cite{yu2018coherent,david2009tissue}, micropipette aspiration \cite{guevorkian2010aspiration} and drop sensors \cite{campas2014quantifying,serwane2017vivo}. In all cases, an external force is used to probe the system. A few methods have been used to obtain relative measurements at the tissue scale such as laser ablation \cite{bonnet2012mechanical} or the fusion of tissue aggregates \cite{jakab2008relating,dechriste2018viscoelastic,david2014tissue}. In both cases the measured velocities can be related to material properties. In the first case, the strain rate is related to ratio of tissue stress $\sigma$ and viscosity $\eta$ \cite{bonnet2012mechanical}, while in the second case the speed of fusion is dictated by the viscocapillary velocity $\gamma/\eta$ \cite{frenkel1945viscous,pokluda1997modification,bellehumeur1998,flenner2012kinetic,dechriste2018viscoelastic}. Of all the previous methods, limited appreciation has been given to the fusion method \cite{stirbat2013fine,gordon1972rheological, david2014tissue}, which is arguably one of the simplest methods to obtain relative measures. Additional advantages of the method are the fact that there is no need of a calibrated probe and it is a non-contact method \cite{sugimura2016measuring}. The fusion of viscoelastic droplets is known to exhibit a phenomenon known as arrested coalescence \cite{dahiya2016arrested,dahiya2017arrested,pawar2012arrested,xie2019geometry}, whereby the degree of coalescence is related to the elasticity of the material. The stable anisotropic shapes it can produce, have been exploited extensively to produce emulsions in a wide range of industries like food, cosmetics, petroleum and pharmaceutical formulations \cite{dahiya2017arrested,muguet2001formulation,garabedian2001model, dahiya2016arrested,pawar2012arrested}. Interestingly, this phenomenon has also been observed in various active matter systems such as ant \cite{hu2016entangled} or bacterial \cite{ponisch2018pili} aggregate colonies. Despite the fact that the sintering of viscous droplets is a well known problem \cite{frenkel1945viscous,pokluda1997modification,bellehumeur1998,flenner2012kinetic,dechriste2018viscoelastic,aid2017predictive,joseph2013fluid}, a mathematical formulation akin to the one of Frenkel and Eshelby \cite{frenkel1945viscous,eshelby1949discussion} for the case of two coalescing viscoelastic solid drops is, to our knowledge, still missing. Furthermore, this phenomenon has received only limited appreciation in the context of tissue engineering and it is only recently that it was reported in biological tissues \cite{tsai2015compaction}. \\

In this work, we report the observation of arrested coalescence in stem cell aggregates and show that a minimal Kelvin-Voigt model successfully captures the dynamics of the process. By fitting our model to the fusion dynamics, the viscocapillary velocity $v_c=\gamma/\eta$ and the shear elastocapillary length $\ell_e=\gamma/\mu$ \cite{bico2018elastocapillarity} can be obtained. In addition, we complement these results with nanoindentation measurements to obtain absoulte values of the effective viscosity, shear modulus and surface tension of the aggregates. Finally, by using agent-based simulations of the fusion process, we propose a mechanism by which active cell protrusions drive a solid-to-fluid phase transition and explore how the supracellular mechanical properties arise from the cell level interactions. \\
\\

\begin{figure}[ht]
\centering
\includegraphics[scale=0.32]{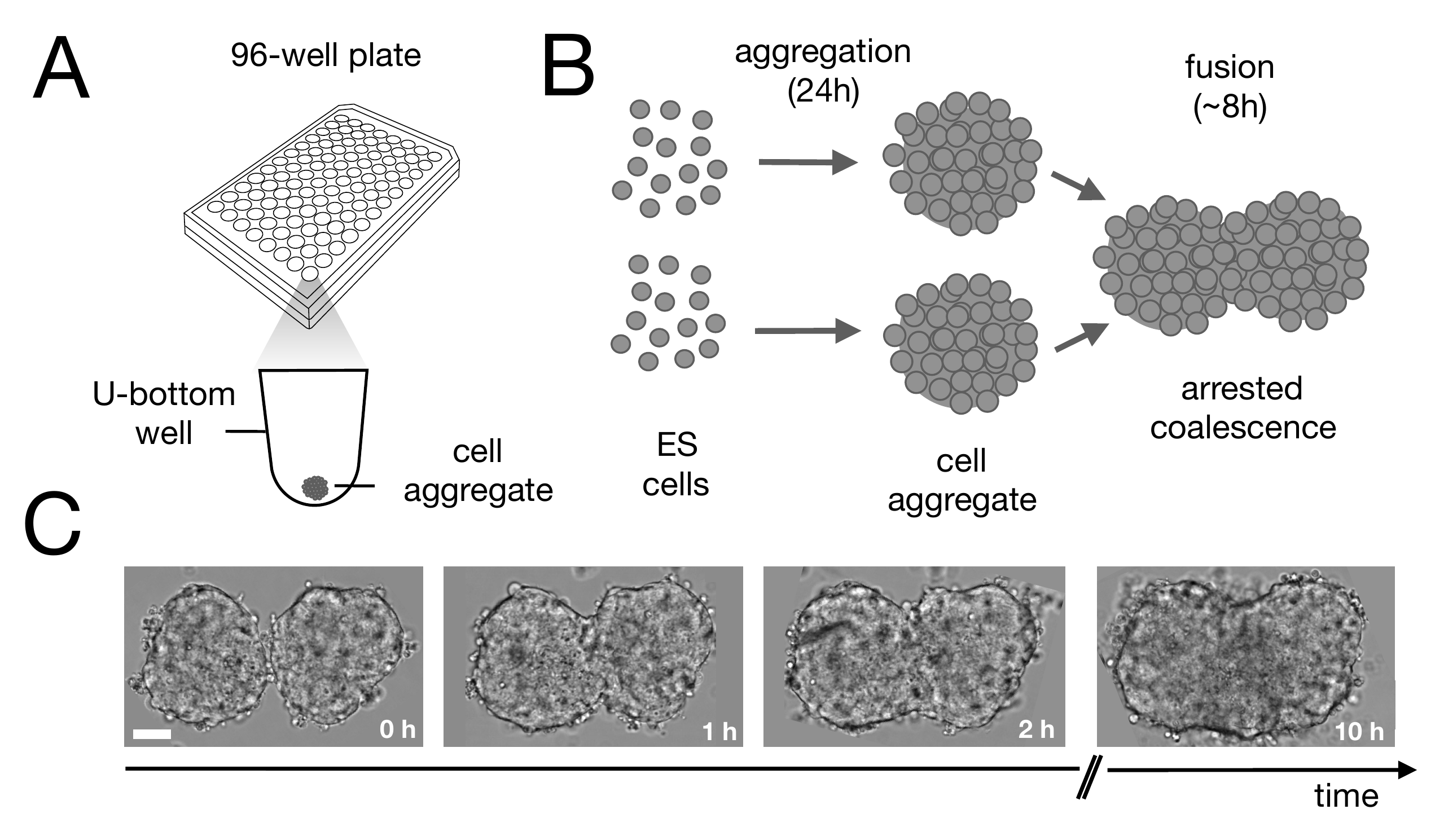}
\caption{Arrested coalescence in aggregates of mouse embryonic stem cells. A) Spheroids were formed by aggregating embryonic stem cells in low adhesion U-bottom multiwell plates. B) Cell aggregates were placed in close contact at 24h after aggregation and the fusion process was imaged using bright field microscopy. C) Image sequence of a fusion event showing the resulting anisotropic shape of the assembly. Notice that the anisotropic shape of the assembly does not change significantly from $\sim 4$h to 10h. Scale bar: 50 $\mu$m. }
\label{fig1}
\end{figure}

{\it Experimental results}. Fusion experiments were carried out by placing two cellular aggregates in contact with each other (Fig. \ref{fig1}A,B). In Fig. \ref{fig1}C an example of the fusion of two aggregates of mouse embryonic stem cells is shown.   Arrested coalescence was observed after $\sim 4$h (Fig. \ref{fig1}C and Movie S1), with anisotropic shapes maintained for the next $\sim 6$h. During the fusion process, the aggregates increased in size due to cell proliferation. To quantify the change in radius we imaged the growth of single aggregates. The radii of the aggregates increased linearly over time. After $\sim 4$h, the radius of the aggregates increased by $\simeq 5 \%$, corresponding to a $\simeq 15$ \% increase in volume (Fig. S1). The doubling time of cells was estimated by simply fitting a linear function to the time evolution of the aggregate radius (see Supplementary Material) and was found to be $T=13.8 \pm 0.4 $ h ($n=10$, mean $\pm$ SD). Given that the fusion process is $\sim 3$ times faster than cell division, we conclude that the volume of the cell aggregates does not change significantly  during the fusion process. \\ 

{\it Continuum model}. In order to understand the fusion dynamics, we considered each multicellular aggregate as a drop of an homogeneous incompressible Kelvin-Voigt material with effective shear viscosity $\eta$, shear modulus $\mu$ and surface tension $\gamma$. This simple constitutive model has been shown to be successful in describing the mechanical properties of multiple types of tissue explants \cite{yu2018coherent} and it is arguably the simplest model to describe arrested coalescence \cite{pawar2012arrested}. The constitutive equation for the stress tensor $\bm{\upsigma}$ reads $\bm{\upsigma}= 2\eta \dot{\bm{\upvarepsilon}} + 2 \mu \bm{\upvarepsilon} - P \mathbb{I}$, where $\bm{\upvarepsilon}= \frac{1}{2} [ \nabla \mathbf{u}+ (\nabla \mathbf{u})^{\text{T}}]$ is the symmetric strain tensor, $P$ is the hydrostatic pressure and $\mathbf{u}$ is the displacement field. Given that cell proliferation is negligible on the timescale of fusion, we approximate the continuity equation as $\nabla \cdot \dot{\mathbf{u}}= 0$. Force balance in the bulk and on the surface read $\nabla \cdot \bm{\upsigma} = 0$ and $\bm{\upsigma} \cdot \mathbf{n} = 2\gamma H \mathbf{n}$, respectively, where $H$ is the local mean curvature of the surface and $\mathbf{n}$ is the unit normal vector to the surface. Next, following the work in Refs. \cite{frenkel1945viscous,pokluda1997modification,bellehumeur1998,flenner2012kinetic}, we approximate the assembly as two identical spherical caps of radius $R(\theta)$ with circular contact `neck' region of radius $r(\theta) = R(\theta) \sin \theta$, with fusion angle $\theta$ (Fig. \ref{fig2}A). The dependence of the radius $R$ on $\theta$ is determined by the incompressibility condition (see Supplementary Material). The dynamics of the fusion process will be described by the evolution of $\theta(t)$. Let us assume the axis of fusion as $\mathbf{e}_x$ (Fig. \ref{fig2}A). The end-to-end length $L(\theta)$ of the fusion assembly along this axis will be given by $L(\theta)=2R(\theta)(1+\cos \theta)$. It is known that coalescence of viscoelastic solid drops can be suppressed for sufficiently large values of the elastic modulus \cite{pawar2012arrested}. The physics at the onset of fusion is not captured by our hydrodynamic model and has its origin on the cell-cell interactions between the two aggregates. To account for such effect we incorporate a pre-strain to the assembly by considering a shift on the rest length $L'(0)=L(0)+\delta L$, being $\delta L/L(0) \ll 1$. The strain is approximated as $\partial_x u \simeq -\varepsilon(\theta)$, with $\varepsilon(\theta) =[L'(0)-L(\theta)]/L'(0) \simeq \varepsilon_P+\varepsilon_L(\theta)$, where $\varepsilon_P=\delta L/L'(0)$ is a pre-strain and $\varepsilon_L(\theta)  = 1- \frac{R(\theta)}{2R_0}(1+\cos \theta)$ is the strain caused by fusion \cite{pawar2012arrested}. The corresponding strain rate reads $\partial_x \dot{u} \simeq - \dot{\varepsilon}(\theta) = \frac{1}{2R_0}\frac{d}{dt} \left[R(\theta)(1+\cos \theta) \right]$. The previous expression differs from the one used in Refs. \cite{frenkel1945viscous,pokluda1997modification,bellehumeur1998,flenner2012kinetic,dechriste2018viscoelastic}, where strain is defined using the distance between the center of a droplet in the assembly and the fusion plane (i.e. $R(\theta) \cos \theta$), as opposed to the end-to-end length $L(\theta)$. Both expressions are only equivalent for small angles (i.e. $\theta \ll 1$). We will use the end-to-end distance definition to be consistent with previous studies on arrested coalescence \cite{pawar2012arrested}, where the maximum strain for complete coalescence reads $\varepsilon_L(\pi/2) =1-2^{-2/3} \simeq 0.37$. Using the previous  expressions we can calculate the dynamics of $\theta$ by equating the work per unit time done by the bulk and surface forces \cite{flenner2012kinetic} (see Supplementary Material). The equation for the dynamics of the fusion angle $\theta(t)$ reads:
\begin{figure}[ht]
\centering
\includegraphics[scale=0.38]{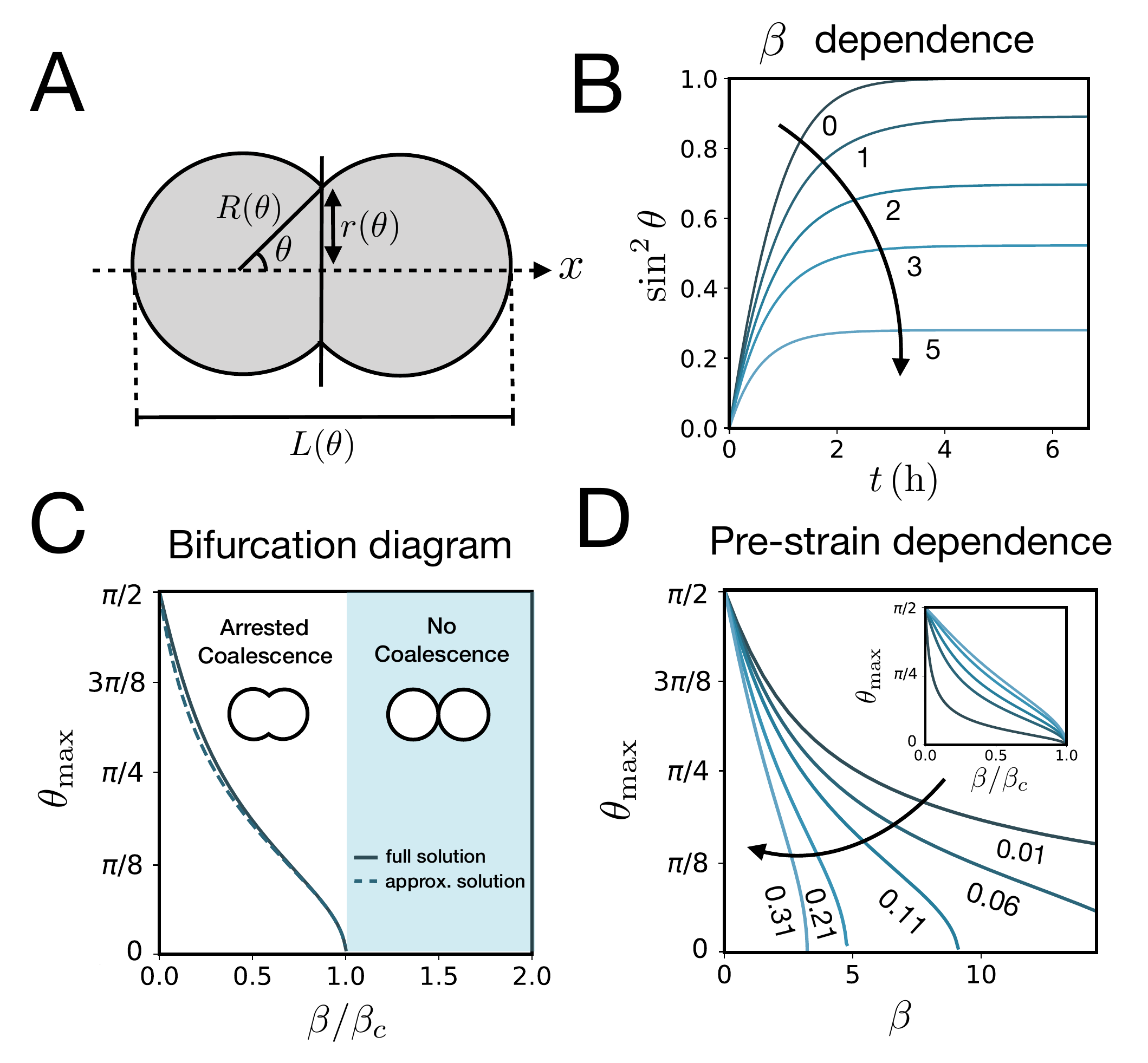}
\caption{A) Schematics of two identical droplets fusing along the $\mathbf{e}_x$ axis. $\theta$ is the angle of fusion which is $\pi/2$ for complete coalescence and takes a value in the range $(0,\pi/2)$ for arrested coalescence. $R(\theta)$ is the radius of each aggregate, $r(\theta)$ is the neck radius during the fusion process and $L(\theta)$ is the end-to-end length. B) Time evolution of $(r/R)^2 = \sin^2 \theta$ as a function of the inverse elastocapillary number $\beta$ for $\tau = 4$ h and $\varepsilon_P = 0.11$ by solving Eq. \ref{Eq1} (see Materials and Methods). C) Bifurcation diagram showing the steady state coalescence angle $\theta_{\rm max}$ as a function of $\beta/\beta_c$. For $\beta<\beta_c$ the system undergoes a pitchfork bifurcation where the non-fused state loses stability in favour of the fused state. The numerical steady state solution of Eq. \ref{Eq1} is shown as a solid line while the approximate analytical solution assuming $R(\theta) \approx R_0$ is shown as a dashed line (see Supplementary Material). $\varepsilon_P = 0.11$. D) Pre-strain dependence of $\theta_{\rm max}$ on $\beta$. Inset: Same plot but rescaling $\beta$ by its critical value $\beta_c$.}
\label{fig2}
\end{figure}
\begin{equation}
\dot{\theta} = \frac{2\cot \theta}{\tau} \left(\frac{R_0}{R(\theta)}\right)^3 [f(\theta)- \beta g(\theta)]
\label{Eq1}
\end{equation}
where $\tau = \eta R_0/\gamma$ is the characteristic viscocapillary time and $\beta = \mu R_0/\gamma$ is a dimensionless parameters characterizing the degree of fusion. The latter dimensionless number is related to the shear elastocapillary length ${\ell_e} =R_0/\beta$. Finally, $f(\theta), g(\theta)$ are functions that depend on the angle $\theta$ (see Supplementary Material). The viscoelastic relaxation time can be obtained as $\tau_v \equiv \eta/\mu =\tau/\beta = \ell_e/v_c$. For small angles and $\beta=0$, Eq. \ref{Eq1} reduces to the typical form for the sintering of viscous drops \cite{bellehumeur1998,flenner2012kinetic}. Considering $\beta$ as our bifurcation parameter, we find that for $\beta > \beta_c= 1/ \varepsilon_P$, elasticity overcomes surface tension and the stable state is $\theta=0$, i.e. no fusion (see Supplementary Material). However, for $\beta < \beta_c$, the system undergoes a pitchfork bifurcation whereby the state $\theta=0$ becomes unstable and droplets fuse (Fig. \ref{fig2},B,C).
This critical condition $\beta=\beta_c$ is equivalent to $\sigma_P = \frac{2 \gamma_c}{R_0}$, which means that coalescence starts when the Laplace pressure equals a pre-stress $\sigma_P=2 \mu \, \varepsilon_P$.  Finally, in Fig. \ref{fig2}D we show the dependence of the maximum fusion angle $\theta_{\rm max}$ on the pre-strain $\varepsilon_P$, which significantly varies for large pre-strains. To fit our model to the experimental data, for simplicity we assumed $\varepsilon_P=0$. Image analysis was performed for each aggregate by using a custom-written software MOrgAna (see Methods). By tracking the end-to-end distance of the assembly $L$ over time, we inferred the time evolution of the fusion angle $\theta$ (see Materials and Methods). The study was carried out by averaging $n=63$ fusion events using different aggregate sizes (see Fig. \ref{fig3}A).  The resulting curves were numerically fitted to the solution of Eq. \ref{Eq1} (Fig. \ref{fig3}A). In Fig. \ref{fig3}B, the inferred parameters $\tau$ and $\beta$ are shown to scale linearly with aggregate size $R_0$, in agreement with our linear viscoelastic solid model. From the slopes of Fig. \ref{fig3}B, we obtain $v_c = 0.09 \pm 0.01$ $\mu$m/min, $\ell_e = 30 \pm 4$ $\mu$m and $\tau_v= 6 \pm 1$ h (mean $\pm$ SE, $n=63$ fusion events). These results were combined with nanoindentation measurements (Fig. S3) where the Hertz model was fitted to the indentation curves (see Methods). The average shear modulus was $\mu = 28 \pm 5$ Pa (mean $\pm$ SE, $n=12$ aggregates), leading to an effective surface tension $\gamma=0.8 \pm 0.2$ mN/m and viscosity $\eta=(5 \pm 1) \cdot 10^5$ Pa$\cdot$s. These values are found within the range of typical values in the literature \cite{forgacs1998viscoelastic, jakab2008relating, guevorkian2010aspiration,stirbat2013fine}. \\

\begin{figure}[ht]
\centering
\includegraphics[scale=0.32]{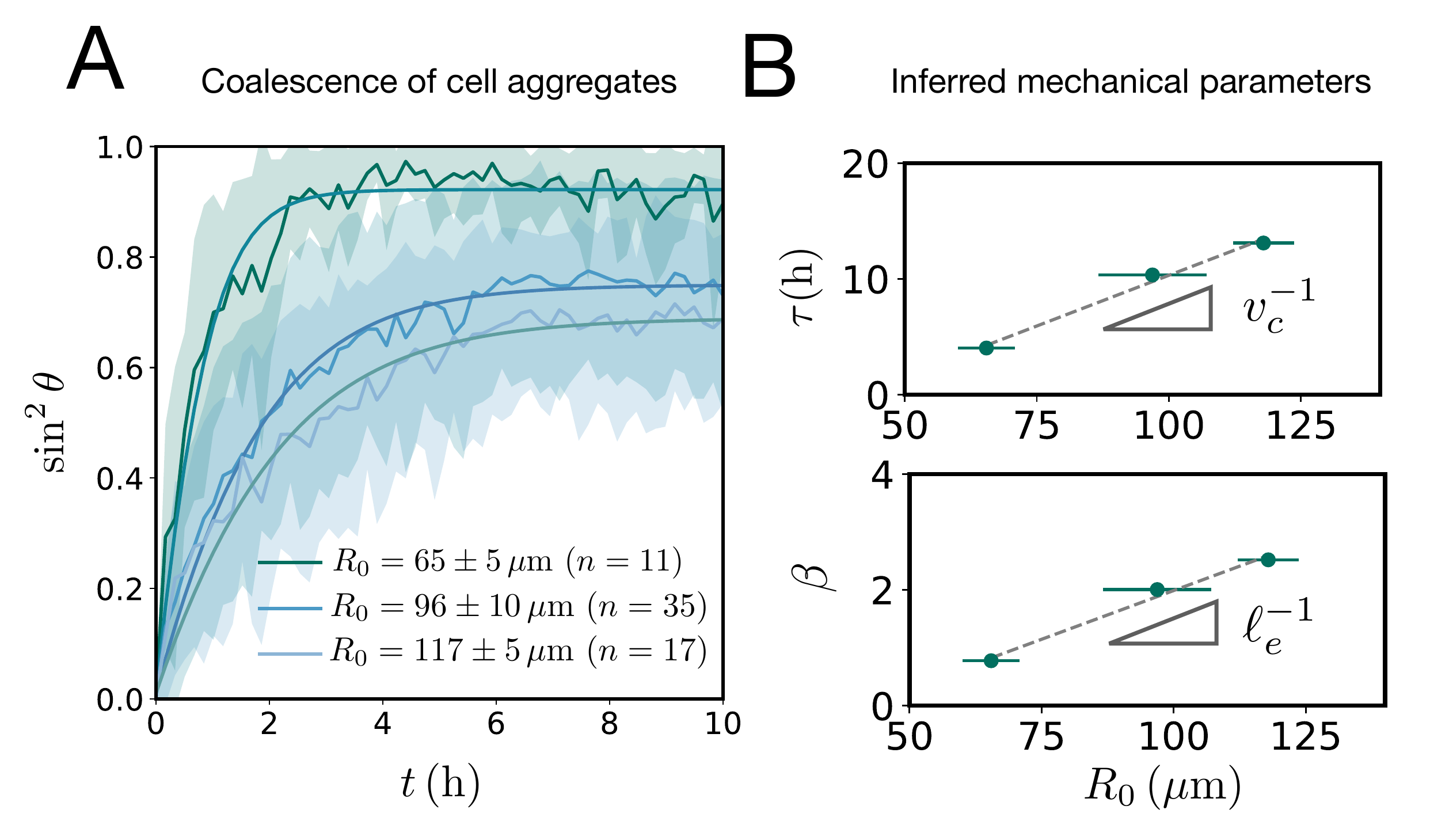}
\caption{A) Fusion dynamics quantification showing the averaged time evolution of $\sin^2 \theta$, where $\theta$ is the fusion angle of the assembly (see Fig. \ref{fig2}). Three different aggregate sizes are shown (shaded regions denote SD around the mean experimental curve). The numerical fits (solid lines) are obtained using Eq. \ref{Eq1}. B) The parameters $\tau$ and $\beta$ scale linearly with aggregate size as expected from the theory. From the slope the viscocapillary velocity $v_c$ and shear elastocapillary length $\ell_e$ can be inferred (mean $\pm$ SD in {\it x}-axis and SE in {\it y}-axis, $n=63$ fusion events). }
\label{fig3}
\end{figure}

{\it Agent-based simulations}. Despite the Kelvin-Voigt model providing a good fit to the experimental data, the rheology of cell aggregates is indeed much more complicated and it is unclear how cell-cell interactions give rise to the observed effective macroscopic mechanical properties. To understand this, we turned to agent-based simulations of cellular aggregates using the GPU-based software ya$||$a (see Fig. \ref{fig3}A), which supports easy implementation of diverse cellular behaviours \cite{germann2019gpu}. For simplicity, we considered a minimal model taking into account adherent and contractile protrusion interactions between cells. The dynamics of a cell $i$ with center at $\mathbf{x}_i$ reads:
\begin{equation}
\lambda \sum_{j} (\dot{\mathbf{x}}_i - \mathbf{\dot{x}}_j) = \sum_{j} (\mathbf{F}^{\rm s}_{ij}+ \mathbf{F}^{\rm a}_{ij})
\end{equation}
where $j$ runs over the nearest neighbours, $\mathbf{F}^{\rm s}_{ij}$ is a passive cell-cell interaction force and $\mathbf{F}^{\rm a}_{ij}$ is an active force to model contractile cell protrusions, which are known to be important in convergence-extension and cell sorting processes \cite{palsson2000model,belmonte2016filopodial}. Friction forces are considered to be proportional to the relative velocity of neighbouring cells with friction coefficient $\lambda$, a typical assumption used in foam and colloidal systems \cite{durian1995foam,tighe2011relaxations} as well as in tissues \cite{mao2013differential, van2015simulating}. Cells have radius $r_0$ and the distance between a pair of cells $i$ and $j$ is denoted as $\mathbf{r}_{ij}=\mathbf{x}_i-\mathbf{x}_j$. The passive cell-cell interaction force consists of two parts: a repulsion harmonic force $\mathbf{F}^{\rm s}_{ij}=K_{\rm r}(2r_0-|\mathbf{r}_{ij}|) \hat{\mathbf{r}}_{ij}$ for $|\mathbf{r}_{ij}|<2r_0$ that describes excluded volume interactions and a truncated harmonic attractive force describing cell-cell adhesion for {$|\mathbf{r}_{ij}|\geq2r_0$ such that $\mathbf{F}^{\rm s}_{ij}=K_{\rm adh}(2r_0-|\mathbf{r}_{ij}|) \Theta(r_{\rm max}-|\mathbf{r}_{ij}|)$, where $\hat{\mathbf{r}}_{ij} = \mathbf{r}_{ij}/|\mathbf{r}_{ij}|$. The active part $\mathbf{F}^{\rm a}_{ij}$ consists of cells randomly selecting a nearest neighbour and applying a constant force $\mathbf{F}^{\rm a}_{ij}=-F_p \hat{\mathbf{r}}_{ij}$ if $|\mathbf{r}_{ij}| \geq 2r_0$, where $F_p>0$ is defined as contractile (see Supplementary Material). The choice of a constant protrusion force is a simplification of more complicated velocity-force relationships found experimentally \cite{heinemann2011keratocyte}. We associate a lifetime to each protrusion $\tau_{\rm on}$ and analogously, a waiting time $\tau_{\rm off}$. Thus a protrusion duty ratio can be defined as $\alpha = \tau_{\rm on}/(\tau_{\rm on}+\tau_{\rm off})$. The protrusion dynamics is similar to a shot noise process of active origin \cite{ben2011effective}. Hence, protrusion interactions introduce force dipoles stochastically in the cell aggregate, which are known to induce cell-cell rearrangements that fluidize tissues \cite{ranft2010fluidization,oriola2017fluidization,petridou2019fluidization}. \\

We analyzed the fusion dynamics in the simulations by using the end-to-end length of the assembly as in the experiments and varied the protrusion force $F_p$ and the duty ratio $\alpha$ (see Fig. \ref{fig3}). We fitted Eq. \ref{Eq1} to the averaged dynamics (Fig. \ref{fig3}B) and extracted the effective macroscopic parameters $\tau$ and $\beta$. The study revealed the presence of three main regimes depending on $\beta$ (see Movies S2-S4): (i) no coalescence ($\beta \gtrsim 20)$, (ii) arrested coalescence ($20 \gtrsim \beta \gtrsim 1 $) and (iii) complete coalescence ($\beta \lesssim 1$) (Fig. \ref{fig3}C), which qualitatively agree with the regimes found in the continuum model (Fig. \ref{fig2}). The same regimes are also identified when studying the characteristic viscocapillary time $\tau$ (see Fig. S2). These results suggest the system undergoes a solid-to-fluid transition for increasing protrusion strength or duty ratio. To assess if the observed transition is similar to a rigidity or a jamming transition, we studied the relative mean squared displacement of cells in our simulations (Fig. \ref{fig3}D). We found that in regimes (i) and (ii) the behaviour was subdiffusive while the behaviour was mainly diffusive in regime (iii). In addition, we observed that the viscoelastic relaxation time $\tau_v$ showed a clear peak at the transition point (see Fig. \ref{fig3}D, inset). 
Finally, by performing compression/relaxation cycles in parallel plate compression simulations on the aggregates (see Fig. S4 and Movies S5, S6) we identified the presence of a yield stress in regime (ii), below which the deformation was not recovered during the relaxation process, indicating a plastic behaviour of the material \cite{balmforth2014yielding}. Hence, we conclude that in our system, arrested coalescence is found at the vicinity of a solid-to-fluid transition, similar to jammed systems \cite{pawar2012arrested}. \\
\begin{figure}[h!]
\centering
\includegraphics[scale=0.27]{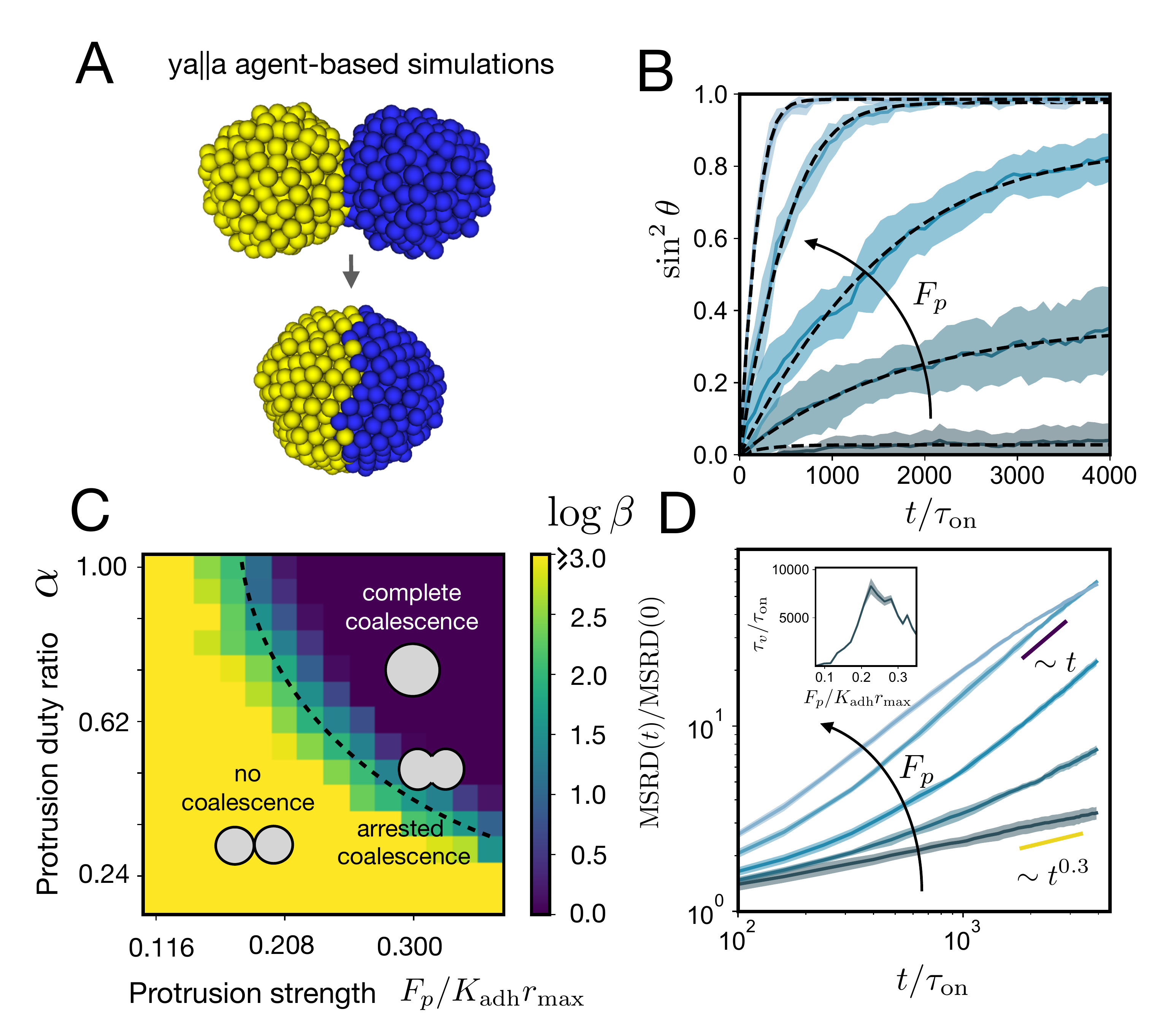}
\caption{A) ya$||$a agent-based simulations of the fusion of two cell aggregates for $F_p/K_{\rm adh}r_{\rm max}=$ 0.20 and $\alpha=1$. B) Averaged time evolution of $\sin^2 \theta$ over time ($n=10$, shaded region denotes SD around the mean simulation curve) in the simulations for different cell protrusion strengths $F_p/K_{\rm adh}r_{\rm max}=$ (0.116, 0.171, 0.208, 0.263, 0.350). $K_{\rm adh}/K_{\rm r}=1$, $r_0/r_{\rm max}=0.4$, $\lambda/K_{\rm adh} \tau_{\rm on}=1$, $\alpha=1$, and $500$ cells per aggregate. The numerical fits (dashed lines) are obtained using Eq. \ref{Eq1}. C) Colormap of $\log \beta$ in parameter space. Three distinct regions can be identified corresponding to no coalescence, arrested coalescence and complete coalescence. D) Mean squared relative displacement of cells as a function of time for different protrusion strengths  $F_p/K_{\rm adh}r_{\rm max}$ (same values as in panel B). Cells change from a subdiffusive ($\sim t^{0.3}$) to a diffusive ($\sim t$) behaviour for increasing $F_p$. Inset: Viscoelastic relaxation time vs protrusion strength.}
\label{fig4}
\end{figure}

{\it Discussion}. Here we report the phenomenon of arrested coalescence in stem cell aggregates and present a viscoelastic theory of sintering to understand the dynamics of the process. We show that a minimal agent-based model considering cell-cell adhesion and dipolar contractile protrusions reproduces arrested coalescence. Additionally, we find that cell protrusion activity controls a solid-to-fluid transition. By combining simulations and continuum theory, we are able to study the dependence of different hydrodynamic quantities on cell-cell interactions. The role of cell protrusions is twofold: on the one hand, it leads to a fluidization process (Fig. \ref{fig3}D) and, on the other hand, it creates an effective surface tension $\gamma$ that drives the coalescence of the cell aggregates. Although the solid-to-fluid transition is of active origin, it is different from other transitions observed in models of self-propelled particles \cite{henkes2011active,bi2016motility} since the protrusion interactions only introduce dipolar forces. The transition we find resembles rigidity transitions recently shown to be driven by active tension fluctuations at cell-cell contacts \cite{kim2020embryonic} or cell-cell adhesions \cite{petridou2021rigidity}. We would like to acknowledge that after our preprint was posted, another group also demonstrated, by means of agent-based simulations, glassy or jammed cell behavior during arrested coalescence of active drops \cite{ongenae2021activity}. Finally, an intrinsic limitation of our particle-based simulations is the absence of cell shape changes which are known to be critical in tissue rheology \cite{manning2010coaction,kim2020embryonic,bi2015density,merkel2018geometrically}. Further work should be done to incorporate such effects for example by means of 3D vertex models \cite{okuda2015three}. \\

Continuum descriptions of drop coalescence have been mainly limited to purely viscous drops \cite{frenkel1945viscous,pokluda1997modification,bellehumeur1998,flenner2012kinetic}. This has limited the use of such theories to the determination of viscosity and surface tension, despite tissue stiffness and viscoelastic effects having important implications for tissue engineering and being known to play a major role in cancer \cite{chaudhuri2020effects,guimaraes2020stiffness}. Here we present a simple method that when combined with a contact method such as nanoindentation or AFM, allows a fast full mechanical characterisation of 3D tissue aggregates. The method successfully described the fusion of human induced pluripotent stem cells (Fig. S5 and Movie S7), thus being a promising tool in the bioengineering and medical fields. More generally, the method can also be potentially used to characterise the mechanics of inert drops in the emulsion industry. Finally, we envision that future work on the theory of sintering for viscoelastic materials will be important in the formation of biological structures {\it in vitro} using bioink units \cite{kosztin2012colloquium}. \\

{\it Acknowledgements}. This work was conceived during the COVID-19 pandemic lockdown. We would like to thank all the healthcare professionals who fought against the disease. All authors were supported by the European Molecular Biology Laboratory (EMBL) Barcelona. We also thank J. Casademunt, X. Diego, M. Merkel, G. Torregrosa and all the members of the Trivedi group for illuminating discussions. D.O. acknowledges funding from Juan de la Cierva Incorporaci\'on with Project no. IJC2018-035298-I, from the Spanish Ministry of Science, Innovation and Universities (MCIU/AEI) and N.G. acknowledges the Human Frontier Science Program (HFSP) with number LT000227/2018-L. Finally, M. M. is supported by the Daiichi Sankyo Foundation of Life Science and Japan Society for the Promotion of Science (JSPS) Overseas Research Fellowships. \\

{\it Author contributions}. Conceptualization: D.O. \& V.T.; Methodology: D.O., M.M.-R., J.S. \& V.T.; Software: N.G., M.M.-R. \& J.S.; Validation: D.O., K.A., G.A., M.M. \& M.M.-R.; Formal analysis: D.O., Investigation: D.O. \& M.M.-R., Resources: M.E., J.S. \& V.T.; Writing: D.O. \& V.T. with inputs from all; Supervision: D.O. \& V.T.; Project administration: V.T.; Funding acquisition: D.O., N.G., M.E., J.S. \& V.T.  \\

\section{Materials and Methods} 

{\bf Mouse ES cell culture and 3D aggregate formation}. Mouse embryonic stem cells (E14 stem cells) were maintained in ES-Lif (ESLIF) medium, consisting of KnockOut Dulbecco's Modified Eagle Medium (DMEM) supplemented with 10\% fetal bovine serum (FBS), 1x Non-essential aminoacids (NEEA), $50$ U/mL Pen/Strep, 1x GlutaMax, 1x Sodium Pyruvate, $50$ $\mu$M 2-Mercaptoethanol and leukemia inhibitory factor (LIF). Cells adhered to 0.1\% gelatin-coated (Millipore, ES-006-B) tissue culture-treated 25 cm$^2$ flasks (T25 flasks, Corning, 353108) in an incubator at 37 $^{\circ}$C and 5\% CO$_2$. To form the aggregates $\sim 300$ cells were aggregated per well in 96-well U-bottom plates (Greiner Cellstar, \#650970) containing $40$ $\mu$L NDiff227 media \, (Takara Bio, \#Y40002) for 24h prior to fusion. \\

{\bf Image acquisition and feature extraction}.  2D images of cell aggregates in 96-well microplates were acquired with the high content imaging PerkinElmer Opera Phenix\textsuperscript \textregistered \, system in non-confocal bright field mode. A 10x air objective was used with 0.3 N.A. and an exposure time of 100 ms. To capture the dynamics of the fusion process, snapshots were acquired every 10 min for a duration of $10$ h. All time points were segmented using the software MOrgAna (Machine-learning based Organoid Analysis), a Python-based machine learning software (\url{https://github.com/LabTrivedi/MOrgAna.git)}. To fit our model to the experiments, the end-to-end distance of the assembly $L$ was obtained by fitting an ellipse to the final mask at every time frame. Finally, using the relationship $L(\theta)=2R(\theta)(1+\cos \theta)$ and considering $L(0)=4R_0$, the time evolution of the fusion angle $\theta(t)$ was obtained (see Fig. \ref{fig3}A). \\

{\bf Fitting procedure}. The experimental and simulated data were fitted to the solution of Eq. \ref{Eq1} using a non-linear least squares method. The solution of Eq. \ref{Eq1} was obtained numerically using the Python solver \texttt{odeint} and the fit was done with \texttt{curve\_fit}, both functions from the Python package \texttt{scipy} \cite{2020SciPy-NMeth}.  \\

{\bf Nanoindentation measurements}. The mechanical measurements were done using the Chiaro Nanoindenter (Optics11) adapted to a Leica DMi8 inverted microscope. The aggregates were transferred from the multiwell plates to $\mu$-Slide 8 well coverslips (ibidi, \#80826) coated with 0.1$\%$ gelatin and containing warm NDiff227. Indentations were done with a spherical cantilever probe of $29.5$ $\mu$m of radius and a stiffness of 0.024 N/m. The approach speed was 5 $\mu$m/s and the contact radius was $\sim 10$ $\mu$m. The effective elastic modulus $E_{\rm eff}$ was calculated by fitting the Hertz's model to the indentation curves and the corresponding shear modulus was obtained as $\mu=(E_{\rm eff}/2)(1-\nu^2)/(1+\nu)$ \cite{field1993simple}, assuming a Poisson's ratio of $\nu=1/2$. The effective elastic modulus for each aggregate was obtained by averaging around 3-4 measurements. Finally, the average effective elastic modulus was obtained by averaging over different aggregates.   \\

{\bf Agent-based simulations}. All agent-based simulations were programmed in CUDA C++ using the ya$||$a modelling framework (\url{https://github.com/germannp/yalla}). The neighbour search method used to determine the pairwise interactions is a modified version of the overlapping spheres method \cite{germann2019gpu, Osborne2017}, where the Gabriel method is applied to a preliminary set of nearest neighbours in order to eliminate neighbour interactions that are being blocked by a third, nearer neighbour \cite{Marin-Riera2015,Delile2017}. The equations of motion (Eq. 2) were solved using the two-step Heun method \cite{germann2019gpu}. For additional details see the Supplementary Material.

\bibliography{fusion_bibliography.bib}
\bibliographystyle{ieeetr}

\end{document}



\newenvironment{sciabstract}{%
\begin{quote} \bf}
{\end{quote}}

\newcommand{\beginsupplement}{%
        \setcounter{table}{0}
        \renewcommand{\thetable}{S\arabic{table}}%
        \setcounter{figure}{0}
        \renewcommand{\thefigure}{S\arabic{figure}}%
        \setcounter{figure}{0}
     }




\title{Arrested coalescence of multicellular aggregates (Supplementary Material)}

\author{David Oriola}\email{david.oriola@embl.es}
\affiliation{ EMBL Barcelona, Dr. Aiguader, 88, 08003, Barcelona, Spain}
\author{Miquel Marin-Riera}
\affiliation{ EMBL Barcelona, Dr. Aiguader, 88, 08003, Barcelona, Spain}
\author{Kerim Anlas}
\affiliation{ EMBL Barcelona, Dr. Aiguader, 88, 08003, Barcelona, Spain}
\author{Nicola Gritti}
\affiliation{ EMBL Barcelona, Dr. Aiguader, 88, 08003, Barcelona, Spain}
\author{Marina Matsumiya}
\affiliation{ EMBL Barcelona, Dr. Aiguader, 88, 08003, Barcelona, Spain}
\author{Germaine Aalderink}
\affiliation{ EMBL Barcelona, Dr. Aiguader, 88, 08003, Barcelona, Spain}
\author{Miki Ebisuya}
\affiliation{ EMBL Barcelona, Dr. Aiguader, 88, 08003, Barcelona, Spain}
\author{James Sharpe} 
\affiliation{ EMBL Barcelona, Dr. Aiguader, 88, 08003, Barcelona, Spain}
\affiliation{Instituci\'o Catalana de Recerca i Estudis Avan\c{c}ats, 08010, Barcelona, Spain}
\author{Vikas Trivedi} \email{trivedi@embl.es} 
\affiliation{ EMBL Barcelona, Dr. Aiguader, 88, 08003, Barcelona, Spain}
\affiliation{EMBL Heidelberg, Developmental Biology Unit, 69117 Heidelberg, Germany}

\date{\today}%
                      
\maketitle

\tableofcontents

\beginsupplement

\section{Cell aggregate growth}

In the experiments, we find that the radius $R$ of a cell aggregate grows linearly with time. This type of growth behaviour has been observed in other avascular multicellular systems such as tumor spheroids \cite{conger1983growth,engelberg2008essential}. A model that recapitulates this type of growth considers that only an outer crust of constant thickness $d$ grows with rate $\Gamma$, while the rest of the spheroid does not proliferate \cite{conger1983growth}. Considering the volume of the spheroid $V$ and the volume of the crust $V_{\rm c}$ we have:
\begin{equation}
\dot{V}= \Gamma V_{\rm c}
\end{equation}
We can rewrite the previous equation in terms of the dynamics of the radius \cite{conger1983growth}:
\begin{equation}
\dot{R} = \frac{ \Gamma}{3} \left[ \frac{3d}{R} - 3\left(\frac{d}{R} \right)^2 + \left( \frac{d}{R} \right)^3 \right] R
\end{equation}
For sufficiently long times $d/R \ll 1$, and hence the dynamics follows $\dot{R} \simeq d \, \Gamma$ which leads to:
\begin{equation}
R(t) \simeq R_0+ d\, \Gamma t
\end{equation}
From the last expression we find that, for constant cell density, the dynamics of the cell number $N(t)$ follow:
\begin{equation}
\frac{N(t)}{N_0} \simeq \left(1+ \frac{d \, \Gamma}{R_0} t \right)^3
\end{equation}
and the doubling time will be:
\begin{equation}
T=\frac{R_0}{d \, \Gamma}(2^{1/3}-1)
\end{equation}

\section{Continuum model}

We consider a multicellular aggregate as a homogeneous incompressible Kelvin-Voigt material drop with effective shear viscosity $\eta$, shear modulus $\mu$ and surface tension $\gamma$. Hereinafter, we use index notation and Einstein's summation convention.  The constitutive equation for the stress tensor $\sigma_{ij}$ reads:
\begin{equation}
\sigma_{ij}= 2\eta \dot{\varepsilon}_{ij}+ 2\mu \, \varepsilon_{ij} - P \delta_{ij}
\label{Eq1}
\end{equation}
where $\varepsilon_{ij}= \frac{1}{2} ( \partial_{i} u_{j}+ \partial_j u_i)$ is the symmetric strain tensor, $P$ is the hydrostatic pressure and $u_i$ is the displacement field. 
The continuity equation reads:
\begin{equation}
\partial_i v_i = 0
\end{equation}
where $v_i=\dot{u}_i$. The latter condition is valid provided that cell proliferation is negligible in the system. Force balance in the bulk in the absence of external forces reads:
\begin{equation}
\partial_j \sigma_{ij} = 0 
\label{Eq2}
\end{equation}
Similarly, force balance on the surface reads:
\begin{equation}
\sigma_{ij} n_{j} = 2\gamma H n_i
\label{Eq3}
\end{equation}
where $H$ is the local mean curvature of the surface and $n_i$ is the unit normal vector to the surface. Let us now consider the fusion of two identical spherical aggregates. The total volume and area of the assembly will be denoted by $V$ and $S$, respectively. The work of the viscoelastic forces per unit time $\dot{W}(t)$ reads:
\begin{equation}
\dot{W}= \int_{\Gamma} \sigma_{ij} \partial_j v_i \, dV
\label{Eq4}
\end{equation}
where $\Gamma(t)$ and $\partial \Gamma(t)$ denote the integration domains of the volume and surface of the assembly, respectively. Using force balance in the bulk (Eq. \ref{Eq2}), the divergence theorem and force balance on the surface (Eq. \ref{Eq3}) one finds \cite{frenkel1945viscous,pokluda1997modification,bellehumeur1998,flenner2012kinetic,dechriste2018viscoelastic}:
\begin{equation}
\int_{\Gamma} \sigma_{ij} \partial_j v_i \, dV = \int_{\Gamma} \partial_j(\sigma_{ij} v_i) dV = \int_{\partial \Gamma} \sigma_{ij} n_j v_i dS = \int_{\partial \Gamma}  2\gamma H n_i v_i \, dS = \dot{W}_{\gamma}
\label{Eq5}
\end{equation}
where the last equality on the right-hand side corresponds to the work done by the surface tension forces per unit time, which we refer as to $\dot{W}_{\gamma}$. At the same time, $\dot{W}_{\gamma}$ can also be expressed simply as:
\begin{equation}
\dot{W}_{\gamma} = - \gamma \frac{dS}{dt}
\label{Eq6}
\end{equation}
Combining Eqs. \ref{Eq4} and \ref{Eq5} we find \cite{frenkel1945viscous,pokluda1997modification,bellehumeur1998,flenner2012kinetic,dechriste2018viscoelastic}: 
\begin{equation}
\dot{W} = \dot{W}_{\gamma}
\label{Eq7}
\end{equation}
The previous expression states that the work done by the bulk forces per unit time equals the work done by the surface forces per unit time. Following previous works \cite{frenkel1945viscous,pokluda1997modification,bellehumeur1998,flenner2012kinetic,dechriste2018viscoelastic,aid2017predictive,joseph2013fluid}, we model the two fusing aggregates as two spherical caps of radius $R(\theta)$ with circular contact `neck' region of radius $r(\theta) = R(\theta) \sin \theta$ (see Fig. 2A in the Main Text). The volume $V(\theta)$ and surface $S(\theta)$ of two fused droplets with fusion angle $\theta$ can be obtained by using simple geometric considerations \cite{dechriste2018viscoelastic}:
\begin{eqnarray}
V(\theta) &=& \frac{2\pi}{3} R^3(\theta) (2-\cos \theta)(1+ \cos \theta)^2 \\
S(\theta) &=& 4 \pi R^2(\theta)(1+\cos \theta)
\label{Eq8}
\end{eqnarray}
Given that $\partial_i v_i =0$, the total volume of the assembly $V$ will be conserved during the fusion process. Considering an initial radius of the aggregates $R(0)=R_0$, the total volume of the assembly reads $V(\theta)=\frac{8}{3} \pi R_0^3$.  From the previous equation we obtain $R(\theta)$ \cite{flenner2012kinetic}:
\begin{equation}
R(\theta)= 2^{2/3}(1+\cos \theta)^{-2/3}(2-\cos \theta)^{-1/3} R_0
\label{Eq9}
\end{equation}
The dynamics of the fusion process will be completely determined by the evolution of $\theta(t)$, with $\theta(0)=0$ to $\theta(\infty) = \theta_{\rm max}$. Let us assume the axis of fusion is $\mathbf{e}_x$ (see Fig. 2A in the Main Text). The end-to-end length $L(\theta)$ of the fusion assembly along this axis will be given by:
\begin{equation}
L(\theta)=2R(\theta)(1+\cos \theta)
\label{Eq10}
\end{equation}
As mentioned in the Main Text, our hydrodynamic model cannot capture the physics at the onset of fusion and we add a pre-strain to effectively account for an elasticity threshold value for the fusion of viscoelastic solid drops \cite{pawar2012arrested}. We consider a shifted rest length $L'(0)=L(0)+\delta L$, being $\delta L/L'(0) \ll 1$.  We approximate the strain as $\partial_x u \simeq -\varepsilon(\theta)$, where $\varepsilon(\theta)$ reads:
\begin{equation}
\varepsilon(\theta) \equiv \frac{L'(0)-L(\theta)}{L'(0)} \simeq \varepsilon_P + \varepsilon_L(\theta)
\label{Eq11}
\end{equation}
with $\varepsilon_P = \delta L/L'(0)$ and $\varepsilon_L(\theta)$ having the following expression \cite{pawar2012arrested}:
\begin{equation}
\varepsilon_L(\theta) = \frac{L(0)-L(\theta)}{L(0)} = 1- \frac{R(\theta)}{2R_0}(1+\cos \theta)
\label{Eq11b}
\end{equation}
We approximate the corresponding strain rate as $\partial_x v \simeq -\dot{\varepsilon}(\theta)$, where $\dot{\varepsilon}(\theta)$ reads:
\begin{equation}
\dot{\varepsilon}(\theta) = - \frac{1}{2R_0}\frac{d}{dt} \left[R(\theta)(1+\cos \theta) \right] 
\label{Eq12}
\end{equation}
The previous expression differs from the usual one, used in Refs. \cite{frenkel1945viscous,pokluda1997modification,bellehumeur1998,flenner2012kinetic,dechriste2018viscoelastic}, which is based on a definition of strain as changes in center-to-center length of the fusion assembly (as opposed to end-to-end as in Eq. \ref{Eq11b}). Using Eq. \ref{Eq11} and the fact that the system is incompressible, we obtain an approximation for the strain tensor:
\begin{equation}
\partial_i u_j \approx -\varepsilon(\theta) 
\begin{pmatrix}
   1& 0 & 0 \\
   0 & -1/2 & 0  \\ 
   0 & 0 & -1/2  \\ 
   \end{pmatrix}
   \label{Eq13}
\end{equation}
Using the previous simplified expressions we can calculate the work per unit time done by the bulk and surface forces:
\begin{eqnarray}
\dot{W}&=& 4 \pi R_0^3 (2\eta \dot{\varepsilon}^2 +E \dot{\varepsilon} \varepsilon) = \frac{\gamma  \pi \dot{\theta}  R^2(\theta) \sin \theta}{2-\cos \theta} \left( \beta \left[-1-\cos \theta + \frac{2R_0}{R}(1+\varepsilon_P) \right] +  \frac{2 \tau \dot{\theta} \sin \theta}{2-\cos \theta}   \right) \\
\dot{W}_{\gamma}&=& - \gamma \frac{dS}{d\theta} \dot{\theta} = 2 \pi \gamma  R^2(\theta)  \left(\frac{ \sin 2 \theta}{2-\cos \theta} \right) \dot{\theta}
\label{Eq14}
\end{eqnarray}
where $\tau = \eta R_0/\gamma$ is the characteristic viscocapillary time and $\beta = \mu R_0/\gamma$ is a dimensionless parameter quantifying the degree of fusion. The shear elastocapillary length reads $\ell_e \equiv \gamma/\mu = R_0/\beta$. Using Eq. \ref{Eq7} we find an equation for the dynamics of the fusion angle $\theta(t)$:
\begin{equation}
\dot{\theta} = \frac{2\cot \theta}{\tau} \left(\frac{R_0}{R(\theta)}\right)^3 [f(\theta)- \beta g(\theta) ]
\label{Eq15}
\end{equation}
where $f(\theta), g(\theta)$ read:
\begin{eqnarray}
f(\theta) &=& \frac{4}{(1+\cos \theta)^2} \\
g(\theta) &=& \frac{2}{\cos \theta (1+ \cos \theta)} \left[ \frac{2R_0(1+\varepsilon_P)}{R(\theta) (1+\cos \theta)} -1\right]
\label{Eq16}
\end{eqnarray}
Notice that from Eq. \ref{Eq15} the viscocapillary time $\tau$ is a factor of 4 larger than the usual definition \cite{frenkel1945viscous,pokluda1997modification,bellehumeur1998,flenner2012kinetic,dechriste2018viscoelastic}. This is a consequence of our choice of strain in Eq. \ref{Eq11b} which is two times smaller than the usual one. For small angles and $\beta=0$, Eq. \ref{Eq15} reduces to the typical form for the sintering of viscous drops \cite{bellehumeur1998,flenner2012kinetic} (see Eq. \ref{Eq20} in {\it Connection to previous studies of viscous drops}). Let us study the stability of the system around $\theta=0$. The dynamics of a perturbation $\delta \theta \ll 1$ reads:
\begin{equation}
\delta \dot{ \theta} = \frac{2(1-\beta \varepsilon_P)}{\tau \delta \theta} + \mathcal{O}(\delta \theta)
\label{Eq16a}
\end{equation}
Hence for $\beta < \beta_c = 1/ \varepsilon_P$, the non-fused state $\theta=0$ loses stability. Notice that in the absence of pre-strain (i.e. $\varepsilon_P=0$) the non-fused state is always unstable and hence, two drops in contact will always fuse. The critical condition is equivalent to:
\begin{equation}
 \sigma_P \equiv 2 \mu \, \varepsilon_P = \frac{2 \gamma_c}{R_0}
 \label{Eq17a}
 \end{equation}
which means that the yield point corresponds to when the pre-stress $\sigma_P$ equals the Laplace pressure. When the Laplace pressure is larger than the pre-stress of the material $\sigma_P> \frac{2 \gamma}{R_0}$, fusion starts.  
Considering $R(\theta) \approx R_0$, we can obtain an analytical expression for $\theta_{\rm max}$ as a function of $\beta$ by solving Eq. \ref{Eq15} at steady state:
\begin{equation}
\theta_{\rm max} \approx \arctan \left[ \frac{2 \sqrt{(\beta_c-\beta)[\beta+ \beta_c(1+\beta )]}}{\beta(2+\beta_c)}\right]
\label{Eq17}
\end{equation}
As expected, the angle of arrested coalescence is independent of the viscocapillary time $\tau$ and only depends on $\beta$. 
Close to the critical point $\beta = \beta_c$, the arrested fusion angle reads $\theta_{\rm max} \sim \epsilon^{1/2}$, where $\epsilon \equiv (\beta_c - \beta)/\beta_c$ is a small parameter that characterizes the distance to the critical point. Hence the system is completely determined by three parameters: $\tau$, $\beta$ and $\varepsilon_P$.
\subsection{Connection to previous studies of viscous drops}
Here we connect our results to previous classical results in the literature of the sintering of purely viscous droplets. For $\beta=\varepsilon_P=0$ and small angles ($\theta \ll1 $), Eq. \ref{Eq15} can be approximated to the well known form:
\begin{equation}
\dot{\theta} \simeq \frac{2 \cot \theta}{\tau}
\label{Eq20}
\end{equation}
This expression is equivalent to the typical form with the only difference being that the viscocapillary time is a factor 4 larger than the usual definition  \cite{kosztin2012colloquium}. By solving this equation we find that $\sin^2 \theta(t)$ follows \cite{kosztin2012colloquium}:
\begin{equation}
\sin^2 \theta(t) \simeq 1-e^{-4t/\tau}
\label{Eq20}
\end{equation}
A well known scaling relation for the evolution of the angle for $t \ll \tau$ is:
\begin{equation}
\theta(t \ll \tau) \simeq 2\sqrt{ \frac{t}{\tau}} \sim t^{1/2}
\label{Eq21}
\end{equation}
Another useful expression is the time dependence of the relative shrinkage of the assembly assuming $R(t) \simeq R_0$:
\begin{equation}
\frac{L(t)}{L(0)} \simeq \frac{1}{2} \left(1+e^{-2t/\tau} \right) 
\end{equation}
the last expression is related to the evolution of the aspect ratio \cite{schoetz2013glassy,brangwynne2011active}. For short-timescales ($t \ll \tau$) the previous expression reduces to:
\begin{equation}
\frac{L(t \ll \tau)}{L(0)} \simeq 1-\frac{t}{\tau} 
\end{equation}

\section{Agent-based simulations}

\subsection{Protrusion dynamics}

 At every time step, the protrusion dynamics is simulated in the following way: for a set of $N$ cells we define $N$ protrusions, each protrusion $P_i$ being produced by each cell $i$. The protrusion $P_i$ always has one end at $\mathbf{x}_i$ and can be connected to another cell $j$ at $\mathbf{x}_j$. At any time point, $P_i$ can be “on”, that is connecting cell $i$ to cell $j$ and applying a force of magnitude $F_p$ (see Main Text), or it can be “off”, that is not connected to a cell $j$ and thus not applying any force. The probabilities that $P_i$ are switched “on” or “off” during the time step $\Delta t$ are given by $\mathcal{P}_{\rm on}= \log 2 \, \Delta t/\tau_{\rm off}$ and $\mathcal{P}_{\rm off}= \log 2 \, \Delta t/\tau_{\rm on}$, respectively. At every time step, for each protrusion $P_i$, it is stochastically determined whether $P_i$ should be updated given the probabilities $\mathcal{P}_{\rm on}$ (if $P_i$ is currently off) or $\mathcal{P}_{\rm off}$ (if $P_i$ is currently on). If that is the case and the current state is “on”, $P_i$ is switched off. Alternatively, if the current state is off, a random cell $\mathbf{x}_j$ is chosen such that it is found at a distance from $\mathbf{x}_i$ smaller than $2r_0$, and $P_i$ is then connected to $\mathbf{x}_j$ at the other end.

\subsection{Fusion and parallel plate compression simulations}

Next, the details for each type of simulation are described: \\

{\it Fusion simulations}: Two separated spherical aggregates are created by randomly generating 3D points within a sphere. Next, we let the system evolve for a short transient to make sure the two aggregates have reached its equilibrium configuration prior to the start of the simulation. Finally, to start the fusion process we move the aggregates closer so that they contact each other. \\

{\it Parallel plate compression simulations}: 

To simulate the process of parallel plate compression we defined the position of the upper/lower plate $z_k$, $k=1,2$ as the position of the uppermost/lowermost cell over time along the $z$-axis. A cell $i$ in the aggregate will experience a force $\mathbf{F}^{\rm c}_{ik}$ from plate $k$ defined as:
\begin{eqnarray}
\mathbf{F}^{\rm c}_{ik} &=&\frac{F}{n_k(t)} \frac{z_{ki}}{|z_{ki}|} \mathbf{e}_z \quad {\rm if } \quad 2r_0-|z_{ki}|>0 \nonumber \\
\mathbf{F}^{\rm c}_{ik} &=& \mathbf{0} \quad {\rm otherwise} 
\end{eqnarray}
where $z_{ki}=z_k-z_i$ is the distance between plate $k$ and a cell $i$ along the $z$-axis, $n_k(t) \in \mathbb{N}$ is the number of cells that fulfil the condition $2r_0-|z_{ki}|>0$ (i.e. interact with plate $k$) at time $t$, and $F$ is the total external force applied to each plate. The extended dynamics of each cell $i$ reads:
\begin{equation}
\lambda \sum_j (\dot{\mathbf{x}}_i - \mathbf{\dot{x}}_j ) = \sum_{j} (\mathbf{F}^{\rm s}_{ij}+ \mathbf{F}^{\rm a}_{ij}) + \sum_{k} \mathbf{F}^{\rm c}_{ik}
\end{equation}
The initial condition consists of a single spherical aggregate of cells and two plates positioned on opposite sides of the aggregate along the Z axis. At the start of the simulation and during a certain time period, an external force of magnitude $F$ is applied in each plate in order to compress the aggregate. After that period, the plates are removed so that relaxation can take place (see Movies S5 and S6). \\

\bibliography{fusion_bibliography.bib}
\bibliographystyle{ieeetr}

\newpage

\begin{figure}[h!]
\centering
\includegraphics[scale=0.68]{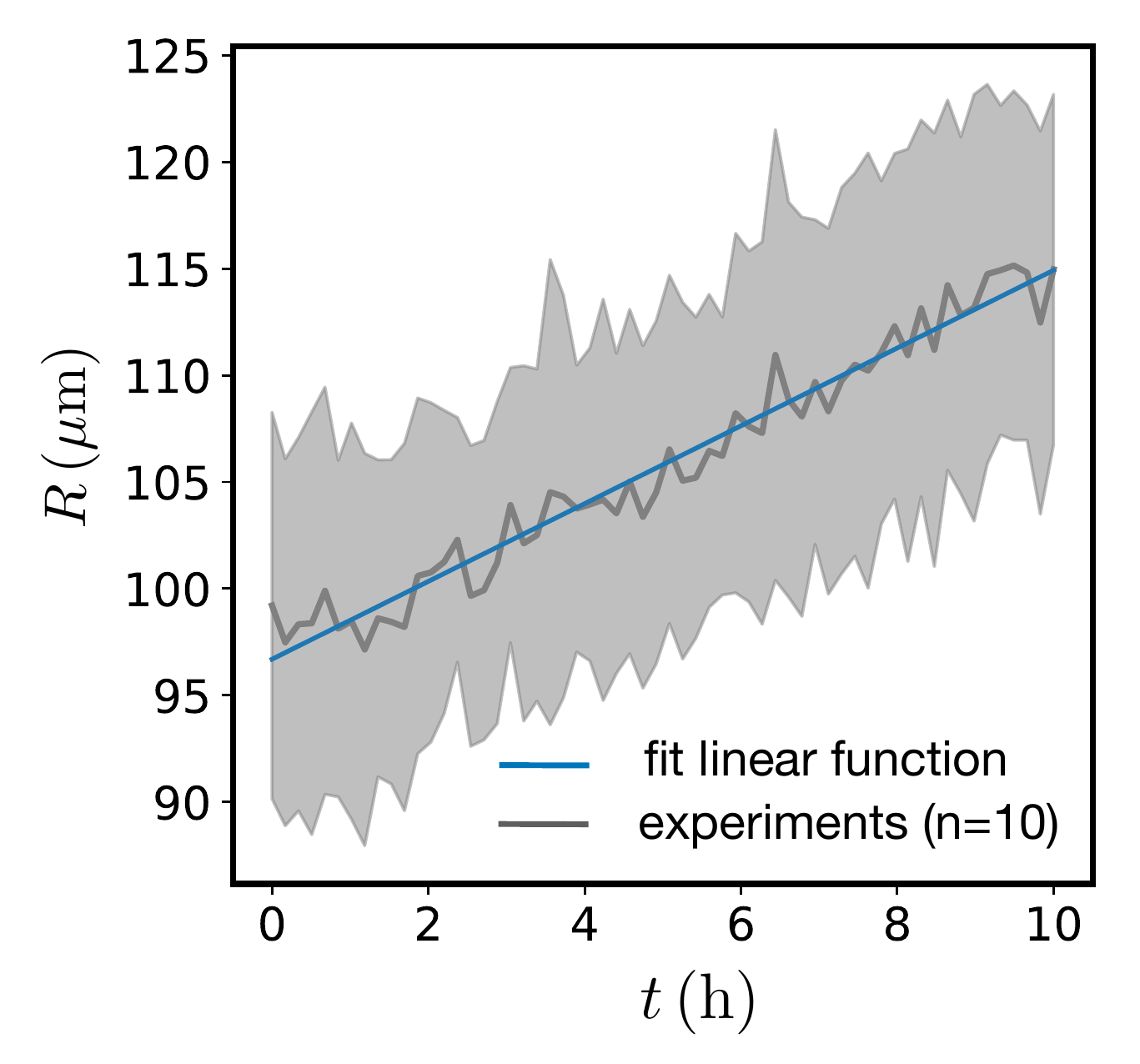}
\caption{Average radius of a single mouse aggregate over time ($n=10$ aggregates, mean $\pm$ SD). The linear fit is used to obtain the doubling time of the cells.}
\label{figS1}
\end{figure}
\begin{figure}[h!]
\centering
\includegraphics[scale=0.60]{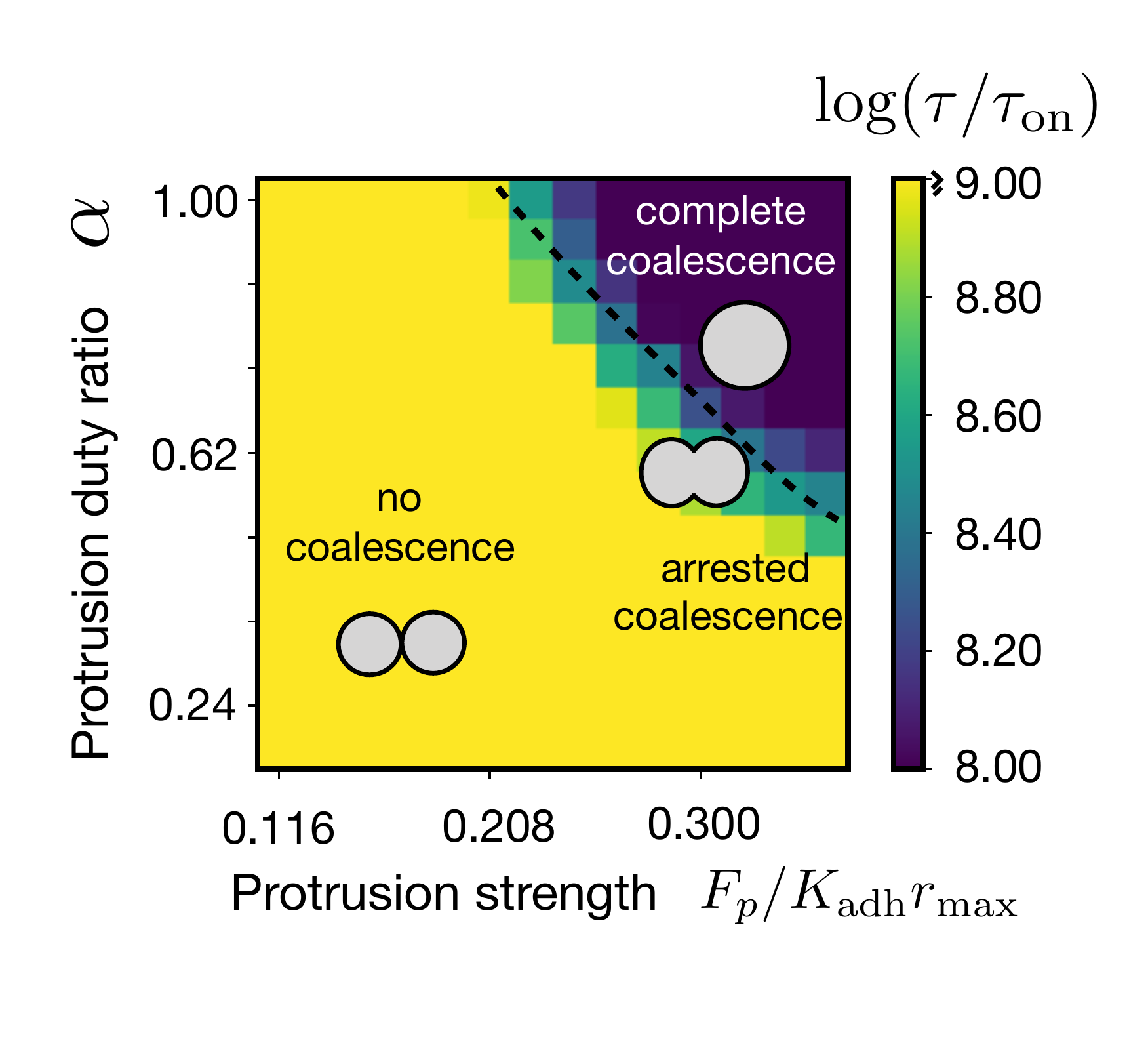}
\caption{Colormap of the logarithm of the dimensionless viscocapillary time $\log(\tau/\tau_{\rm on})$ as a function of the protrusion duty ratio $\alpha$ and the protrusion strength $F_p/(K_{\rm adh} r_{\rm max}$). The parameters are the same as in Fig. 4 in the Main Text.}
\label{figS2}
\end{figure}

\begin{figure}[h!]
\centering
\includegraphics[scale=0.45]{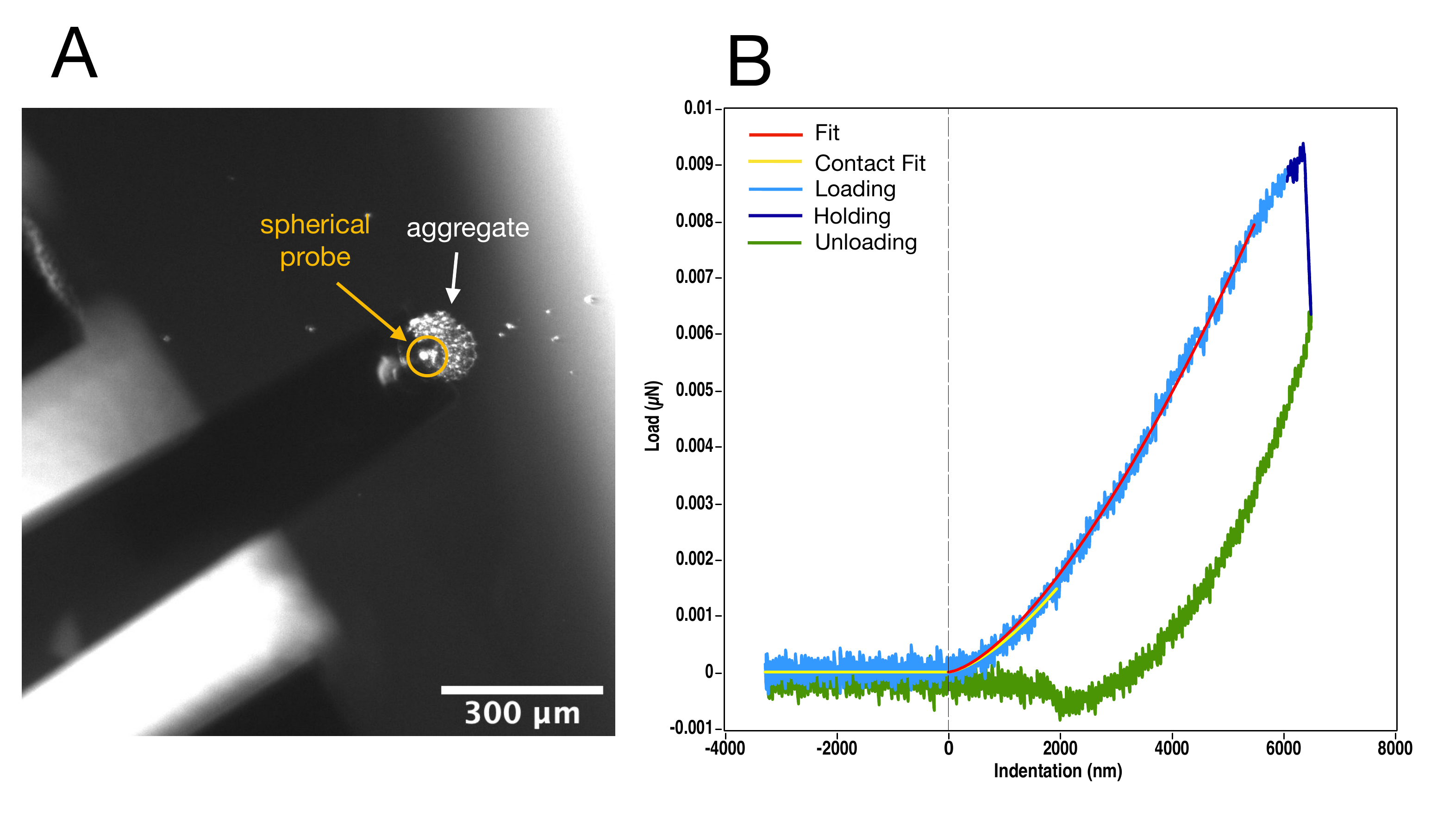}
\caption{Nanoindentation. A) Bright field image of a 24h mouse embryonic stem cell aggregate in NDiff227 medium under the Chiaro Nanoindenter. B) Typical indentation curve with a maximal force of $\sim 0.01$ $\mu$N and maximum indentation depth of $\sim 7$ $\mu$m. The fits correspond to the Herz's model considering a spherical tip. }
\label{figS3}
\end{figure}
\begin{figure}[h!]
\centering
\includegraphics[scale=0.42]{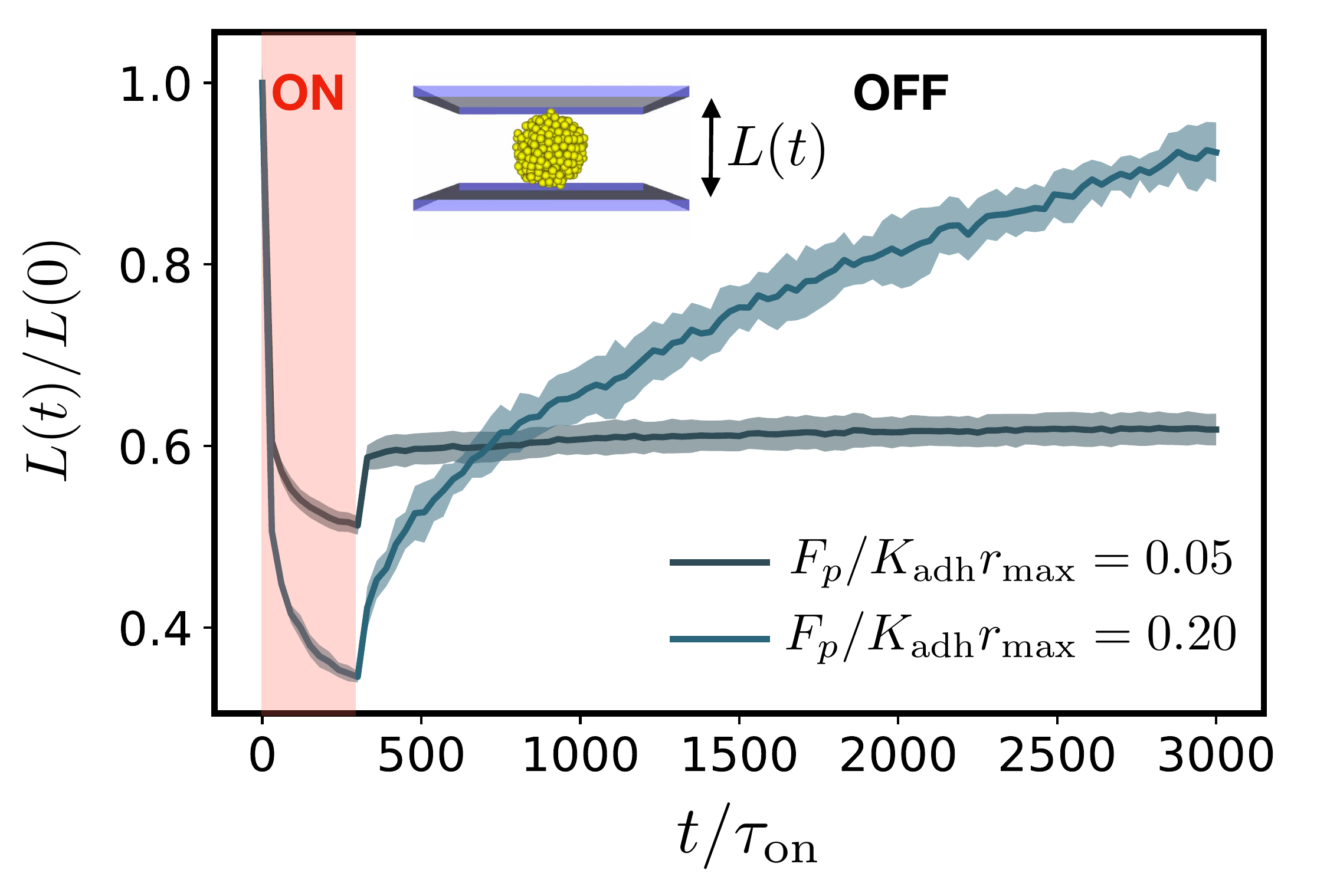}
\caption{Parallel plate compression simulations using the software ya$||$a. The axial deformation of the aggregate $L(t)/L(0)$ (mean $\pm$ SD, $n=9$) is studied considering jammed ($F_p/(K_{\rm adh}r_{\rm max})=0.05$) and unjammed ($F_p/(K_{\rm adh}r_{\rm max})=0.20$) cell aggregates of 500 cells each. $K_{\rm adh}/K_{\rm r}=1$, $\alpha=1$, $\lambda/K_{\rm adh} \tau_{\rm on}=1$. In the red region, a compressive force $F/(K_{\rm adh}r_{\rm max})=10$ is applied. See corresponding Movies S5 and S6.}
\label{figS4}
\end{figure}
\begin{figure}[h!]
\centering
\includegraphics[scale=0.48]{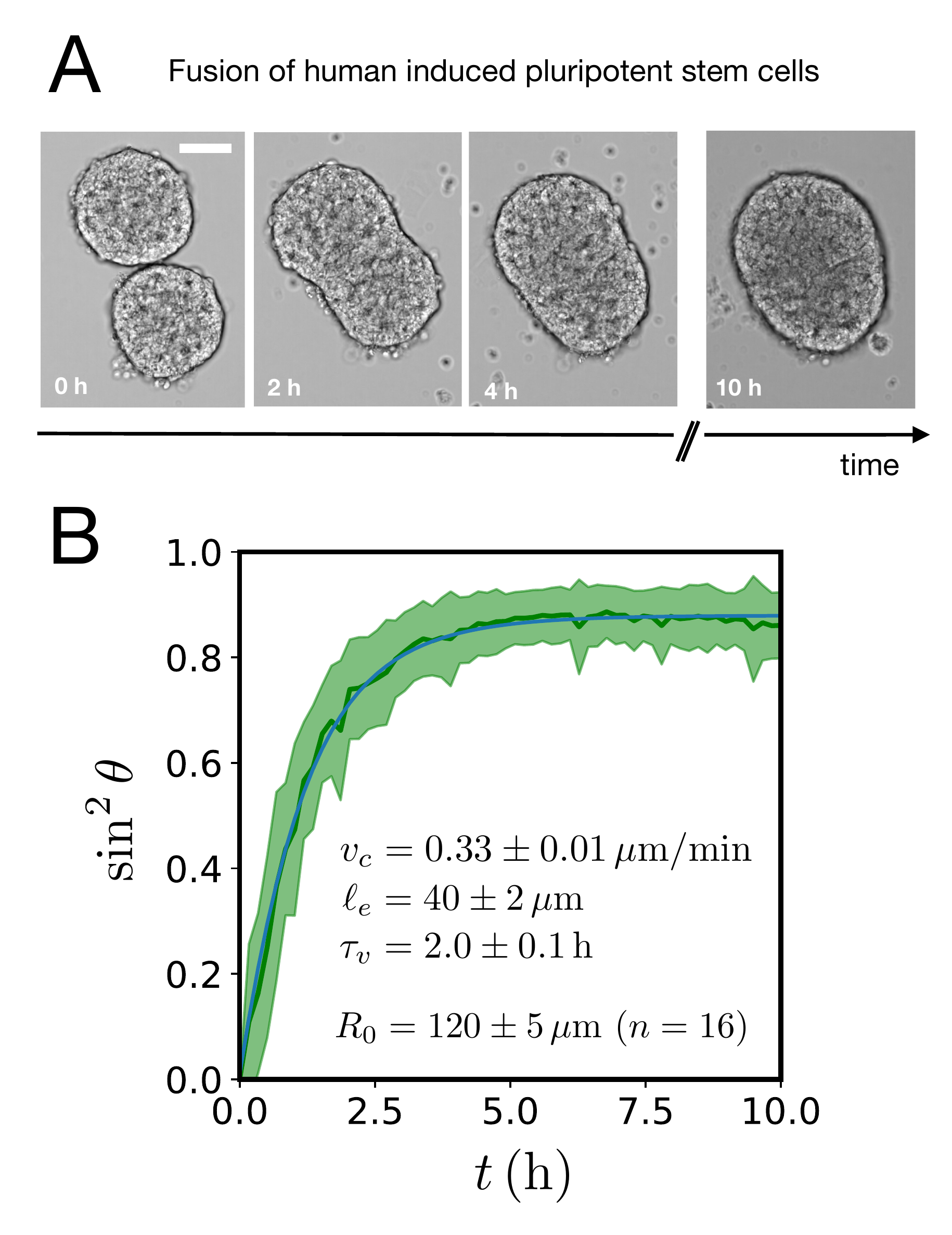}
\caption{Fusion of human induced pluripotent stem cell (hiPSCs) aggregates at 24h after aggregattion. The number of cells per aggregate was $\sim 500$ during aggregation. A) Sequence of snapshots showing a fusion event of hiPSCs. Scale: 100 $\mu$m. B) Average curve of $\sin^2 \theta$ vs time for $n=16$ fusion events with an average radius of $R_0 = 120 \pm 5$ $\mu$m (mean $\pm$ SD). The estimated values of $v_c$, $\ell_e$ and $\tau_v$ are shown.}
\label{figS5}
\end{figure}
\begin{figure*}[h!]
\centering
\includegraphics[scale=0.42]{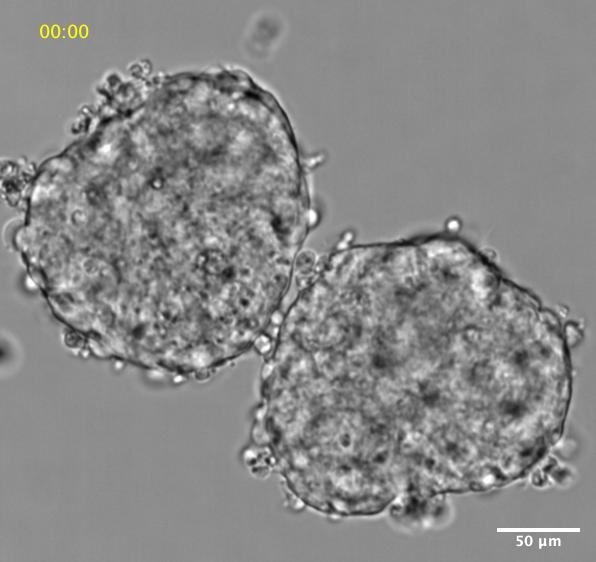}
\caption*{Movie S1. Brightfield timelapse of the fusion of two mouse embryonic stem cell aggregates 24h after aggregation. 10x air objective and 0.3 N.A. The movie corresponds to the snapshots shown in Fig. 1C.}
\label{MovieS1}
\end{figure*}
\begin{figure*}[h!]
\centering
\includegraphics[scale=0.38]{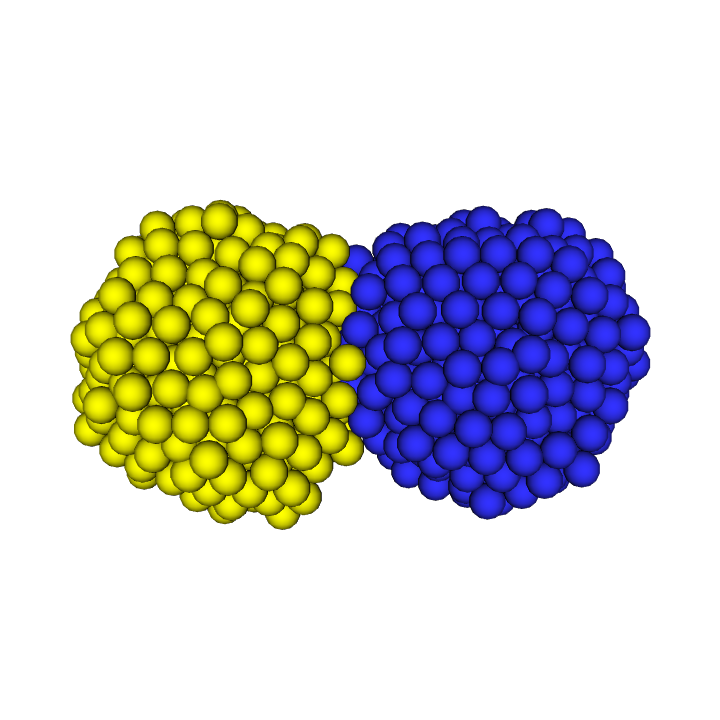}
\caption*{Movie S2. ya$||$a simulation showing no coalescence. 500 cells per aggregate, $K_{\rm adh}/K_{\rm r}=1$, $\alpha=1$, $F_p/(K_{\rm adh}r_{\rm max})=0.10$ and $\lambda/K_{\rm adh} \tau_{\rm on}=1$.}
\label{MovieS2}
\end{figure*}
\newpage
\begin{figure*}[h!]
\centering
\includegraphics[scale=0.38]{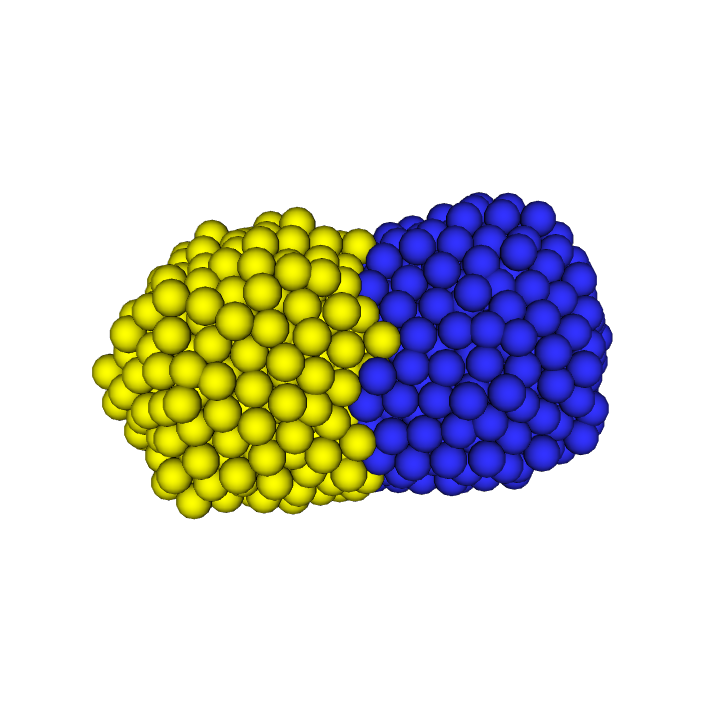}
\caption*{Movie S3. ya$||$a simulation showing arrested coalescence. 500 cells per aggregate, $K_{\rm adh}/K_{\rm r}=1$, $\alpha=1$, $F_p/(K_{\rm adh}r_{\rm max})=0.15$ and $\lambda/K_{\rm adh} \tau_{\rm on}=1$.}
\label{MovieS3}
\end{figure*}
\newpage
\begin{figure*}[h!]
\centering
\includegraphics[scale=0.38]{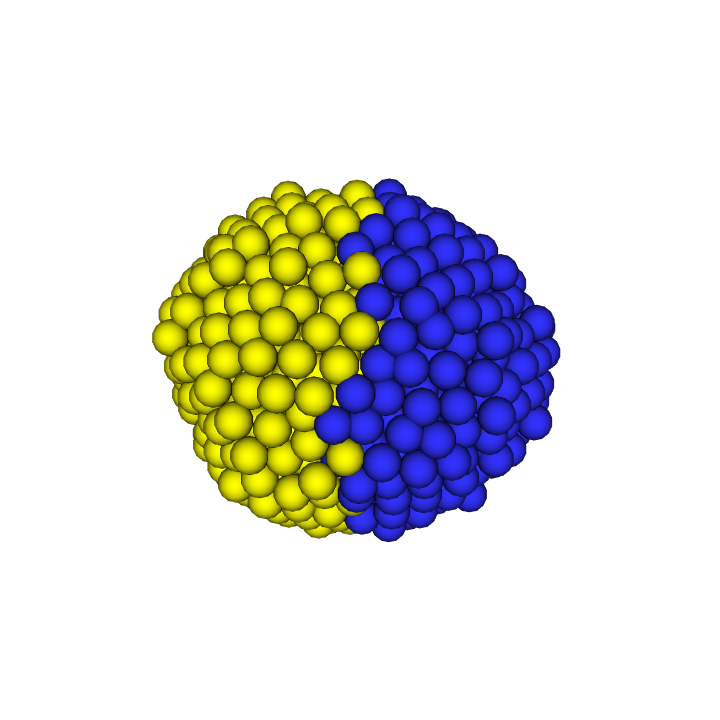}
\caption*{Movie S4. ya$||$a showing complete coalescence. 500 cells per aggregate, $K_{\rm adh}/K_{\rm r}=1$, $\alpha=1$, $F_p/(K_{\rm adh}r_{\rm max})=0.20$ and $\lambda/K_{\rm adh} \tau_{\rm on}=1$.}
\label{MovieS4}
\end{figure*}
\newpage
\begin{figure*}[h!]
\centering
\includegraphics[scale=0.28]{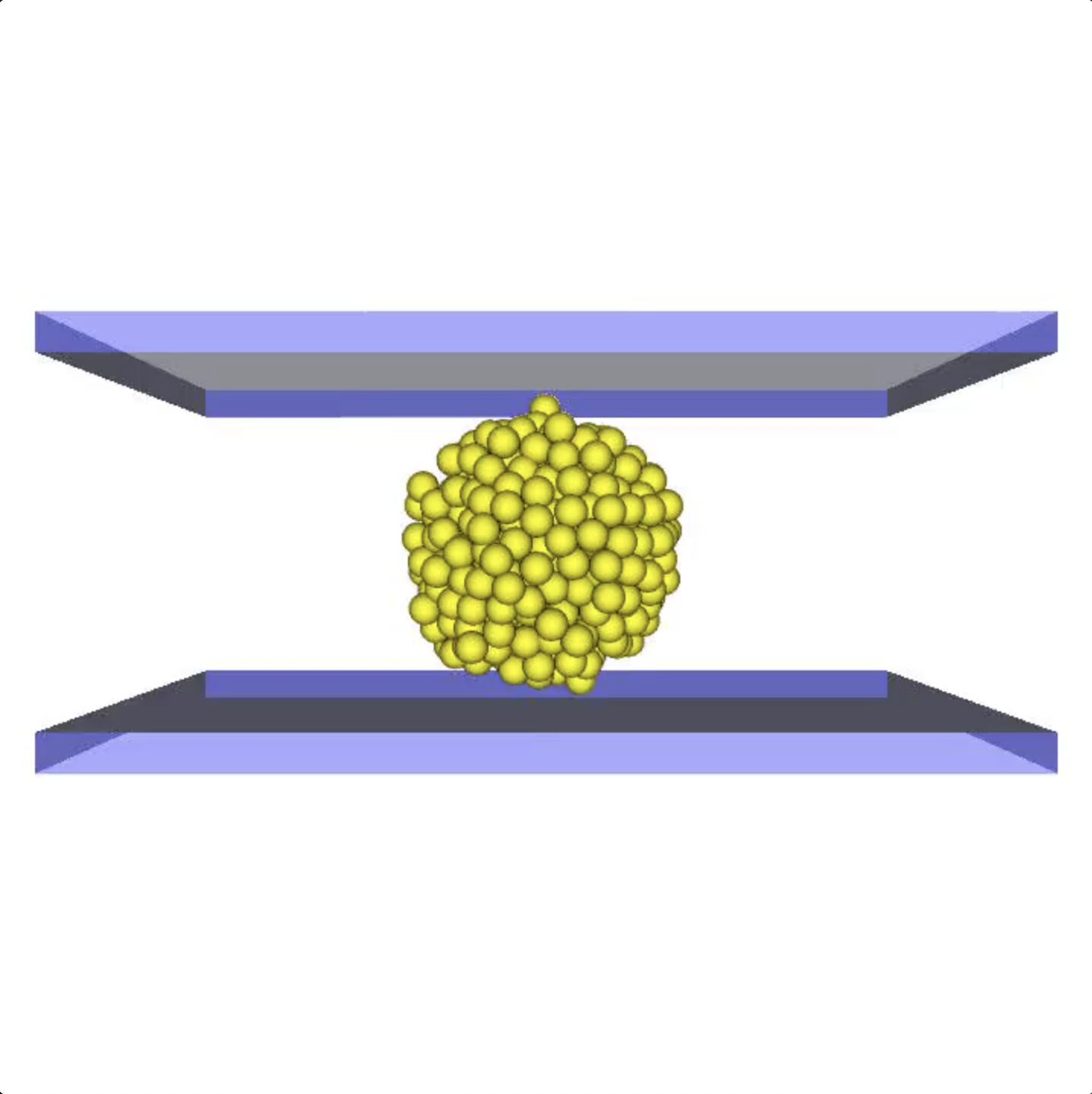}
\caption*{Movie S5. ya$||$a simulation showing no recovery of a 500 cells aggregate in the jammed (solid) regime.  $K_{\rm adh}/K_{\rm r}=1$, $\alpha=1$, $F_p/(K_{\rm adh}r_{\rm max})=0.05$, $F/(K_{\rm adh}r_{\rm max})=10$ and $\lambda/K_{\rm adh} \tau_{\rm on}=1$.}
\label{MovieS5}
\end{figure*}
\newpage
\begin{figure*}[h!]
\centering
\includegraphics[scale=0.28]{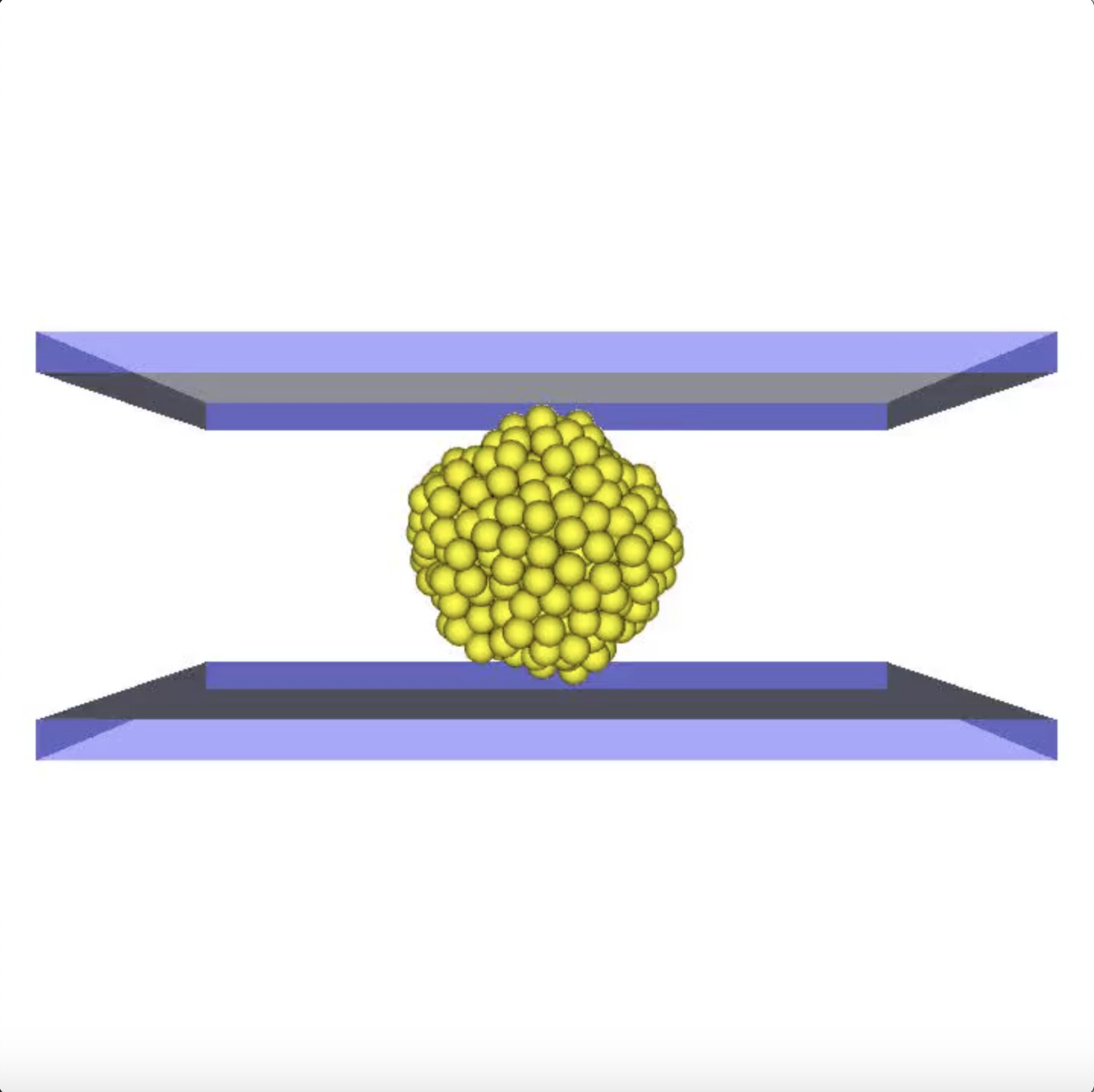}
\caption*{Movie S6. ya$||$a simulation showing complete recovery of a 500 cells aggregate in the unjammed (fluid) regime. $K_{\rm adh}/K_{\rm r}=1$, $\alpha=1$, $F_p/(K_{\rm adh}r_{\rm max})=0.20$, $F/(K_{\rm adh}r_{\rm max})=10$ and $\lambda/K_{\rm adh} \tau_{\rm on}=1$.}
\label{MovieS6}
\end{figure*}
\begin{figure*}[h!]
\centering
\includegraphics[scale=0.30]{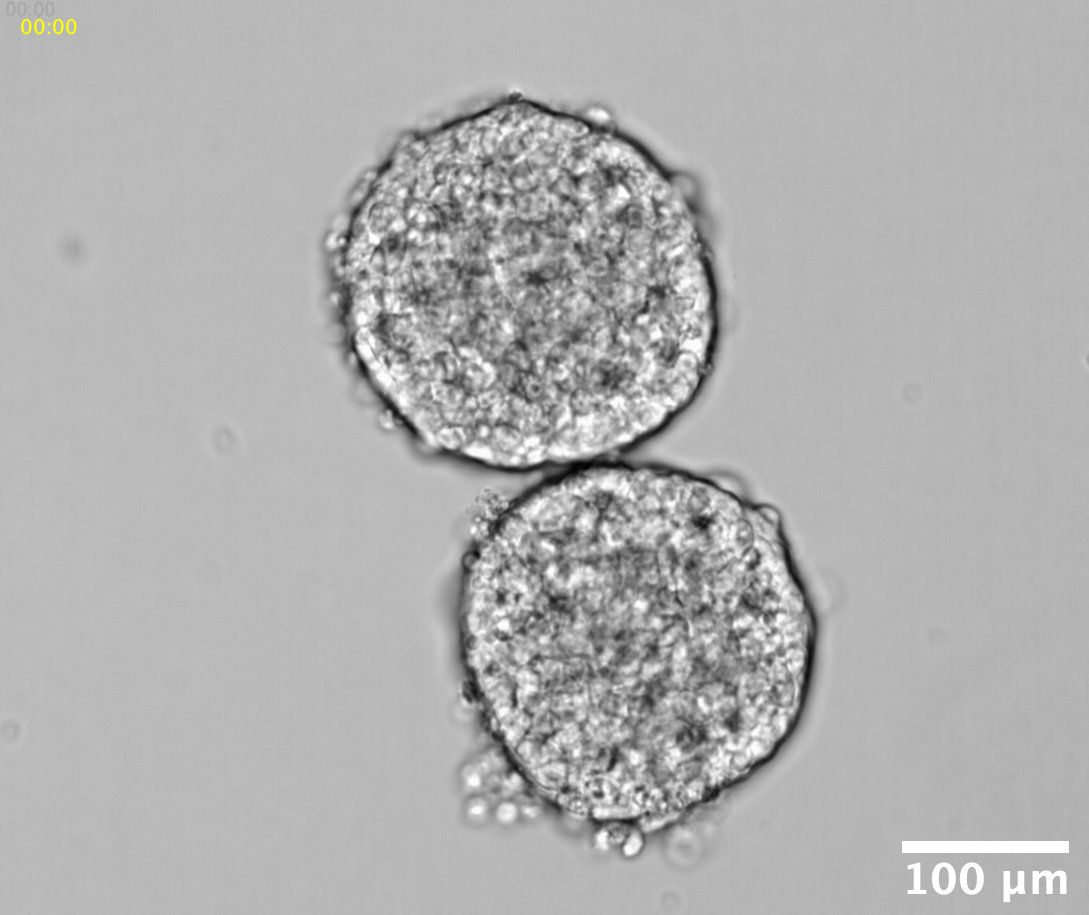}
\caption*{Movie S7. Brightfield timelapse of the fusion of two hiPSC aggregates at 24h. 10x air objective and 0.3 N.A.}
\label{MovieS7}
\end{figure*}